\documentclass{article}
\usepackage{geometry}
\usepackage[utf8]{inputenc}
\usepackage{mathtools}
\usepackage[T2A]{fontenc}
\usepackage[english]{babel}
\usepackage{float}
\usepackage[bottom]{footmisc}
\usepackage{tikz}
\definecolor{universityterracotta}{HTML}{C65A3A}
\usetikzlibrary{arrows.meta}
\usepackage{comment}
\usepackage{xcolor}
\usepackage{amsmath,amssymb,amsfonts,amsthm}
\usepackage{graphicx}
\usepackage{comment}
\newtheorem{prop}{Proposition}[section]
\newtheorem{defi}[prop]{Definition}
\newtheorem{lemma}[prop]{Lemma}
\newtheorem{cor}[prop]{Corollary}
\newtheorem{rem}[prop]{Remark}
\newtheorem{example}[prop]{Example}
\newcommand{\bref}[1]{\textbf{\ref{#1}}}

\newcommand{\dl}[1]{\mathchoice{\ffrac{\partial}{\partial #1}}{\frac{\partial}{\partial #1}}{\ffrac{\partial}{\partial #1}}{\ffrac{\partial}{\partial #1}}}

\newcommand{\ffrac}[2]{\raisebox{.5pt}{\footnotesize$\displaystyle\frac{#1}{#2}$}\kern1pt}
\newcommand{\tensor}{\otimes}
\newcommand{\cN}{\mathcal N}

\numberwithin{equation}{section}
\newcommand\blfootnote[1]{%
  \begingroup
  \renewcommand\thefootnote{}\footnote{#1}%
  \addtocounter{footnote}{-1}%
  \endgroup
}

\newtheorem{defin}{Definition}

\usepackage{amsmath,amssymb}
\usepackage{booktabs}
\usepackage{makecell}

\newcommand{\fR}{\mathbb{R}
}

\DeclareMathOperator{\sgn}{sgn}
\usepackage{hyperref}

\usepackage{amsmath}
\usepackage{array}
\usepackage{tabularx}

\newcolumntype{Y}{>{\centering\arraybackslash}X}
\newcolumntype{Z}[1]{>{\centering\arraybackslash}m{#1}}

\title{Two-scale system: compactified ambient metrics and \\ Scri/Spi/Ti}
\author{Xavier Bekaert\textsuperscript{a}, Maxim Grigoriev\textsuperscript{b,\textdagger}, Mikhail Markov\textsuperscript{b}}
\date{}
\begin{document}

\maketitle
\begin{center}
\small
\textsuperscript{a} Institut Denis Poisson, Université de Tours, Université d’Orléans,
CNRS, IDP, UMR 7013,\\ Parc de Grandmont,
37200 Tours, France

\vspace{2mm}

\textsuperscript{b} Physique de l’Univers, Champs et Gravitation,
Université de Mons -- UMONS,\\
Place du Parc 20,
7000 Mons, Belgium

\vspace{2mm}

\end{center}

\begin{abstract}
We show that certain geometric constructions, including the Fefferman--Graham ambient metric, Brinkmann spaces, and ultrastatic vacuum spacetimes, admit natural conformal compactifications and can be unified within a single framework arising naturally from conformal geometry. The framework is based on a two-scale system comprising two weight-one conformal scalars satisfying almost-Einstein equations and certain regularity conditions. As a particular instance of this unifying framework, we obtain a new ambient construction for Spi and Ti, the extended spatial and temporal infinities of Ricci-flat spacetimes. A remarkable feature of this construction is that Spi/Ti and the null infinity of the corresponding spacetimes are naturally realised as a boundary face and the corner of the two-scale ambient space, respectively. We also demonstrate how the two-scale system can serve as a background ambient geometry for studying the asymptotic behaviour of (gauge) fields at spatial, temporal, and null infinity.
\end{abstract}
\vfill
\blfootnote{%
Email: \texttt{xavier.bekaert@univ-tours.fr},
\texttt{grigoriev.max@gmail.com},
\texttt{markovmikhailv@gmail.com}.
\par
\textsuperscript{\textdagger} Also at Lebedev Physical Institute
and Institute for Theoretical and Mathematical Physics,
Lomonosov MSU.
}

\thispagestyle{empty}

\pagebreak

\tableofcontents

\thispagestyle{empty}

\pagebreak

\pagenumbering{arabic} 

\setcounter{page}{1}

%%%%%%%%%%%%%%%%%%%%%%
\section{Introduction}
%%%%%%%%%%%%%%%%%%%%%%
The idea that a certain (typically more complicated) geometric structure on a $d$-dimensional manifold can be lifted to a (typically simpler) structure on a higher-dimensional manifold and then studied using the tools associated with the latter appears repeatedly in mathematics and theoretical physics. A well-known example is Dirac's observation \cite{Dirac:1936fq} that conformally invariant equations on four-dimensional spacetime can be studied through their lift to a flat six-dimensional ambient space. A curved version of this idea in arbitrary dimension is provided by the Fefferman--Graham ambient construction \cite{Fefferman-Graham:1985ambient,FeffermanGraham2012}. An alternative interpretation of the same construction is that it provides the embedding of AdS (Einstein) space into the flat (resp. Ricci-flat) ambient space of one dimension higher.  Another  realisation of the same principle is the Bargmann structure (sometimes called Eisenhart-Duval lift) \cite{Duval:1984cj,Duval:1990hj}, which encodes $(d+1)$-dimensional Newton--Cartan geometry in terms of $(d+2)$-dimensional Lorentzian geometry.

A striking feature of the Dirac--Fefferman--Graham ambient space construction is that it provides a useful tool to access the conformal infinity of the asymptotically AdS (AAdS) spaces or more precisely it gives a specific explicit realisation of its conformal compactification. Namely the conformal  boundary of AAdS space is realised as a submanifold of the ambient space, more precisely, as a section of the  (curved) null hypercone. Remarkably, in the ambient approach this submanifold is not a conformal boundary of the AAdS space  seen as an embedded curved hyperboloid but is still conformally isometric to it, at least locally. This gives a powerful tool to study the asymptotic boundary structure (boundary values) of the AAdS gravitational field itself and the (gauge) fields on its background. Although the idea was somewhat implicitly present already in the Dirac and the Fefferman-Graham constructions (see also~\cite{Angelopoulos:1980wg,Ferrara:1998bv}), the explicit implementation was given in~\cite{Bekaert:2012vt,Bekaert:2013zya} in the case of gauge fields on AdS space while the curved version and its higher-spin generalisation was discussed in~\cite{Bekaert:2017bpy}.

In this work, we show that the Fefferman--Graham construction, Bargmann structures and ultrastatic vacuum spacetimes can be unified within a single system, called the two-scale sytstem,  that arises naturally from the perspective of conformal geometry. As a new result, we also obtain from this unifying framework an analogue of the ambient construction for spatial and temporal infinities of Ricci-flat spacetimes. Moreover, the conformal compactification of this ambient construction simultaneously provides access to the null infinity of the Ricci-flat spacetime. We also employ the two-scale system to develop an associated ambient formulation of a massive scalar field and show how its asymptotic behaviour at spatial and temporal infinity can be naturally interpreted from the ambient perspective. Let us stress that in contrast to AAdS space the spatial/temporal infinity of asymptotically flat space is not a conformal boundary and hence is not easily accessible using the usual conformal geometry methods, see however~\cite{Borthwick:2023lye,Borthwick:2024wfn}.

To make the discussion more concrete we introduce the following:
\begin{defin}
A two-scale system is a system of equations on a $(d+2)$-dimensional manifold $M$ (possibly with boundary or corner) for a conformal metric $G$ of signature $(p+1,q+1)$, with $p+q=d$, and two weight-one Weyl densities $\Omega_1,\Omega_2$, called scales:
\begin{align}\label{two-scales}
\begin{aligned}
&\nabla_A\nabla_B\Omega_i
+P_{AB}\Omega_i
+G_{AB}\rho_i
=0\,,
\\
&\Omega_i\rho_j+\Omega_j\rho_i
+G^{AB}\partial_A\Omega_i\,\partial_B\Omega_j
=\lambda_{ij}\,,
\end{aligned}
\qquad i,j=1,2\,.
\end{align}
where $\lambda_{ij}$ are the entries of a constant symmetric $2\times 2$ matrix $\lambda=(\lambda_{ij})$, $\nabla_A$ and $P_{AB}$ denote the Levi-Civita connection and the Schouten tensor associated with $G$. Note that $\rho_i$ are auxiliary fields uniquely determined by the first equations and given explicitly by
\begin{align}
\label{rhoi}
    \rho_i=-\frac{1}{d+2}(\nabla^2\Omega_i+P\Omega_i)\,,
\end{align}
where $\nabla^2:= G^{AB}\nabla_A\nabla_B$ and $P:= P_A{}^{A}$.
\end{defin}
\noindent By a conformal metric, we mean a metric tensor of Weyl weight two. Throughout the paper, the indices $i,j$ take two values and label the two scales, while uppercase Latin indices $A,B,\ldots$ range over $d+2$ values. In applications, we mainly consider the Lorentzian signature $(p+1,q+1)=(1,d+1)$ or the conformal signature $(p+1,q+1)=(2,d)$.\footnote{We use the mostly minus signature convention throughout.} We adopt the following definition of the Schouten tensor:
\begin{align}
    P_{AB}=\frac{1}{d}\big(R_{AB}-\frac{R}{2(d+1)}G_{AB}\big)\,.
\end{align}
The two-scale system is Weyl-invariant. In particular, the first line of \eqref{two-scales} consists of two almost-Einstein equations (see, e.g., \cite{BEG,CurryGover2018}), while the second line fixes the Gram matrix of the tractors formed by $(\Omega_i,\nabla_A \Omega_i,\rho_i)$. Let us note that multiple almost-Einstein structures, mostly in Riemannian signature, have been studied in the context of tractor geometry; see~\cite{gover2010class,leitner2012multiple}.

The second line of \eqref{two-scales} almost follows from the first one. Indeed, it is a standard property of the almost-Einstein equation that its differential prolongation implies $\partial_A\rho_i=P_{A}{}^{B}\partial_B\Omega_i$; see, e.g., Section 3.4 of \cite{CurryGover2018}. A straightforward computation then gives
\begin{align}
    \partial_C(\Omega_i\rho_j+\Omega_j\rho_i
+G^{AB}\partial_A\Omega_i\partial_B\Omega_j)=0\,.
\end{align}
Thus, the role of the second line in \eqref{two-scales} is merely to fix the integration constants $\lambda_{ij}$. However, as we shall see below, it is precisely the choice of $\lambda_{ij}$ that distinguishes between the different underlying geometries.

An interesting interpretation of the above two-scale system arises from its relation to a question posed by Brinkmann~\cite{Brinkmann:1925fr} about a century ago: which Einstein metrics can be Weyl-transformed into other Einstein metrics? Equivalently, which conformal classes admit two distinct Einstein representatives? Indeed, the two-scale system describes precisely this situation, since each almost-Einstein equation in \eqref{two-scales} singles out an Einstein metric in the conformal class of $G$ wherever the corresponding scale is nonzero. The nontrivial case is, of course, when the two metrics are not related by a constant rescaling, as will be ensured by the regularity conditions imposed below.

The common zero locus of the two scales, denoted by
\begin{align}\label{corner}
    \mathcal{C}:=\{\Omega_1=\Omega_2=0\}\,,
\end{align}
is of particular interest in what follows. We assume\footnote{Notice that these conditions cannot be satisfied in every signature for any choice of $\lambda$. For example, if $\lambda_{11}=0$, the second equation in \eqref{two-scales} gives $(G^{AB}\partial_A\Omega_1\partial_B\Omega_1)|_{\Omega_1=0}=0$. In Riemannian signature, this implies $d\Omega_1\big|_{\Omega_1=0}=0$, violating both regularity conditions.  Nevertheless, we shall assume that the matrix $\lambda$ and the signature are chosen so that these compatibility conditions can be satisfied if needed.} the following regularity conditions:
\begin{align}\label{regularity}
     \Omega_1d\Omega_2-\Omega_2d\Omega_1\big|_{M\setminus\mathcal{C}}\neq0\,,\qquad d\Omega_1\wedge d\Omega_2\big|_{\mathcal{C}}\neq0\,.
\end{align}
These conditions exclude degenerate zero loci of $\Omega_i$, ensuring that they are regular hypersurfaces rather than arbitrary closed subsets. Note also that, in a sufficiently small neighbourhood of $\mathcal{C}$, the first condition in \eqref{regularity} follows from the second one.
We do not address global issues in this paper and therefore restrict our attention to a neighbourhood of $\mathcal{C}$.

%\vspace{3mm}

The two-scale system defined above is the main object of study in the present work. We show that several constructions arising in theoretical physics, both known and, to the best of our knowledge, previously unnoticed ones, are described by this system for different choices of the matrix $\lambda$. They are summarised in Tables \bref{tab:systems} and \bref{tab:systems2}.

Let us stress that, for different choices of the matrix $\lambda$, restricting solutions of the two-scale system to the domain $\Omega_1\neq 0$ gives rise (say, in Lorentzian signature) to several well-known structures: the Fefferman--Graham ambient construction, ultrastatic vacuum spacetimes, a subclass of Einstein warped products, and Bargmann structures. These correspond to the four inequivalent choices of the matrix $\lambda$. In this sense, these various structures are unified within a single framework. However, requiring the solutions to extend to $\mathcal{C}$ while satisfying the regularity conditions \eqref{regularity} leads to richer geometric structures, as detailed in Tables \bref{tab:systems} and \bref{tab:systems2}.

%\newpage

\begin{table}[!htb]
\centering

\renewcommand{\tabularxcolumn}[1]{m{#1}}
\setlength{\tabcolsep}{2.5pt}
\renewcommand{\arraystretch}{1.9}

\newcommand{\headercell}[1]{%
  \begin{minipage}[c][4.6em][c]{\linewidth}
    \centering
    #1
  \end{minipage}%
}

\newcommand{\tblmatrix}[4]{%
  \begingroup
  \renewcommand{\arraystretch}{1.1}%
  \begin{pmatrix}
    #1 & #2\\
    #3 & #4
  \end{pmatrix}%
  \endgroup
}

\begin{tabularx}{\linewidth}{
    >{\hsize=1.20\hsize
      \linewidth=\hsize
      \centering\arraybackslash}X
    |
    >{\centering\arraybackslash}m{2.05cm}
    |
    >{\hsize=1.05\hsize
      \linewidth=\hsize
      \centering\arraybackslash}X
    |
    >{\centering\arraybackslash}m{2.55cm}
    |
    >{\centering\arraybackslash}m{2.40cm}
    |
    >{\hsize=0.75\hsize
      \linewidth=\hsize
      \centering\arraybackslash}X
}
\hline

\headercell{Two-scale\\[2pt]system}
&
\headercell{Matrix $\lambda$}
&
\headercell{Wedge\\[2pt]
  \mbox{$\Omega_1\Omega_2\neq0$}}
&
\headercell{Level surface\\[2pt]
  \mbox{$\Omega_1=\Omega_2\neq0$}}
&
\headercell{Face\\[2pt]
  \mbox{$\Omega_2=0,\;\Omega_1\neq0$}}
&
\headercell{Corner\\[2pt]
  \mbox{$\Omega_i=0$}}
\\
\hline

Compactified Fefferman-Graham system
&
$\tblmatrix{0}{-1}{-1}{0}$
&
Ricci-flat ambient space
&
Einstein hypersurface
&
Curved null cone
&
Conformal space
\\
\hline

Compactified Spi/Ti ambient system
&
$\tblmatrix{\pm1}{0}{0}{0}$
&
Einstein ambient space
&
Ricci-flat hypersurface
&
Spi/Ti
&
Scri
\\
\hline
\end{tabularx}

\caption{Two inequivalent cases of two-scale systems with a wedge structure,
together with their geometrical interpretation. The three columns with
$\Omega_1\neq0$ are interpreted in terms of the ambient metric
$\tilde G=\Omega_1^{-2}G$.}
\label{tab:systems}

\vspace{1cm}

\begin{tabularx}{\linewidth}{
    >{\hsize=1.07\hsize
      \linewidth=\hsize
      \centering\arraybackslash}X
    |
    >{\centering\arraybackslash}m{2.05cm}
    |
    >{\hsize=0.93\hsize
      \linewidth=\hsize
      \centering\arraybackslash}X
    |
    >{\centering\arraybackslash}m{2.55cm}
    |
    >{\centering\arraybackslash}m{2.40cm}
}
\hline

\headercell{Two-scale\\[2pt]system}
&
\headercell{Matrix $\lambda$}
&
\headercell{Wedge\\[2pt]
  \mbox{$\Omega_1\Omega_2\neq0$}}
&
\headercell{Level surface\\[2pt]
  \mbox{$\Omega_1=\Omega_2\neq0$}}
&
\headercell{Codimension-two surface\\[2pt]
  \mbox{$\Omega_i=0$}}
\\
\hline

Compactified Einstein degression system
&
$\tblmatrix{\pm1}{0}{0}{\pm1}$
&
Einstein\qquad\qquad ambient space
&
Einstein hypersurface
&
Conformal space
\\
\hline

Compactified Brinkmann system
&
$\tblmatrix{0}{0}{0}{0}$
&
Ricci-flat Brinkmann space
&
Carrollian hypersurface
&
Doubly-degenerate conformal space
\\
\hline
\end{tabularx}

\caption{Two inequivalent cases of two-scale systems without boundary face or
corner, and their geometrical interpretation. The two columns with
$\Omega_1\neq0$ are interpreted in terms of the ambient metric
$\tilde G=\Omega_1^{-2}G$.}
\label{tab:systems2}

\end{table}
\newpage

The plan of the paper is as follows. Section \bref{generalanalysis} provides a general analysis of the two-scale system and introduces all the necessary material for the rest of the paper (such as important definitions and useful coordinate systems) so that, from Section \bref{generalanalysis}, the reader can jump to any of the following four sections (i.e. Sections \bref{FGsection}-\bref{CompactifiedBrinkmannspace}) where we examine, one by one, the four types of inequivalent two-scale systems.
We start with the compactified Fefferman-Graham system in Section \bref{FGsection}. Then we define and discuss the compactified Spi/Ti ambient system in Section \bref{sec:ads-ambient} where we describe how to embed the extended spatial/temporal infinity of the induced metrics on the level hypersurface as a null submanifold inside the ambient space. The third type of two-scale system, the compactified Einstein degression, is briefly discussed in Section \bref{Einsteinsingularfoliation}. The compactified Brinkmann system is treated in Section \bref{CompactifiedBrinkmannspace}, where we establish its relation to Bargmann structures. Lastly, in Section~\bref{sec: fields}, we briefly discuss fields on a two-scale background, taking a massive Klein-Gordon field as an example. 

\paragraph{Note added:} While this paper was under completion, we learned from J.~Borthwick and Y.~Herfray about their paper in preparation \cite{Borthwick:2026} which appears to have some overlap with the present one.

%%%%%%%%%%%%%%%%%%%%%%%%%%%%%%%%%%%%%%%%%%%%%%%%%%
\section{General analysis of the two-scale system}\label{generalanalysis}
%%%%%%%%%%%%%%%%%%%%%%%%%%%%%%%%%%%%%%%%%%%%%%%%%%

We begin by discussing the rigid symmetries of the two-scale system. We then discuss the equivalent ``gauge-fixed'' form of the two-scale system obtained by imposing the additional condition $\Omega_1\neq0$. We will also obtain some general results on two-scale systems, such as the form of the metric in adapted coordinates.

%%%%%%%%%%%
\subsection{Symmetries and redefinitions}\label{sec:symmetries}
%%%%%%%%%%%

Any two-scale system \eqref{two-scales} is manifestly invariant under the following gauge symmetries: diffeomorphisms and Weyl transformations. However, it is also manifestly invariant under some rigid symmetries.

\begin{prop}\label{prop:stabilizer}
Consider the group
\begin{align}
    Stab(\lambda)
    =
    \{A\in GL(2,\mathbb{R})\mid A\lambda A^{T}=\lambda\}\,.
\end{align}
Then, for any $A\in Stab(\lambda)$, the transformation
\begin{align}\label{GLtransfos}
    \Omega_i \mapsto \Omega'_i=A_i{}^j\,\Omega_j
\end{align}
preserves equations \eqref{two-scales}.
\end{prop}
\noindent In particular, if $\lambda_{11}=\lambda_{22}$, the system is invariant under a $\mathbb{Z}_2$ symmetry
\begin{align}
\Omega_1 \leftrightarrow \Omega_2.
\end{align}

Notice that when $A\in GL(2,\mathbb{R})$ but does not belong to the stabilizer of $\lambda$, the transformation
\begin{align}
    \Omega_i'=A_{i}{}^{j}\Omega_j\,,\qquad \lambda'=A\lambda A^T
\end{align}
is not a symmetry of the two-scale system associated with \(\lambda\). Nevertheless, it defines an invertible field redefinition that establishes a one-to-one correspondence between the solutions of the systems associated with $\lambda$ and $\lambda'$. Consequently, two matrices belonging to the same $GL(2,\mathbb R)$-congruence orbit, i.e. $\lambda'=A\lambda A^T$, define equivalent two-scale systems, whereas matrices belonging to different orbits define genuinely inequivalent systems. Representatives of these orbits are listed in Table~\bref{tab:stabilizer-orbits}.

\begin{table}[ht]
\centering
\renewcommand{\arraystretch}{1.3}
\begin{tabular}{c|c|c|c}
\hline
Two-scale system 
&
Representative 
&
Stabiliser $Stab(\lambda)$
&
Orbits of $Stab(\lambda)$
\\
&
of the matrix $\lambda$
&
of the matrix $\lambda$
&
in the  $(\Omega_1,\Omega_2)$-plane
\\
\hline
\begin{tabular}{c}
Compactified \\ Einstein \\degression
\end{tabular}
&
$
\begin{pmatrix}\pm1
&0\\0&\pm1
\end{pmatrix}$
&
$O(2)$
&
\begin{tabular}{c}
$\Omega_1=\Omega_2=0$;\\
$\Omega_1^2+\Omega_2^2=r^2>0$
\end{tabular}
\\
\hline
\begin{tabular}{c}
Compactified\\
Fefferman-Graham\\
system
\end{tabular}
&
$\begin{pmatrix}0&-1\\-1&0\end{pmatrix}$
&
$O(1,1)$
&
\begin{tabular}{c}
$\Omega_1=\Omega_2=0$;\\
$\Omega_1\Omega_2=s>0$;\\
$\Omega_1\Omega_2=s<0$;\\
$\Omega_1\Omega_2=0,\;
(\Omega_1,\Omega_2)\neq(0,0)$
\end{tabular}
\\
\hline
\begin{tabular}{c}
Compactified Spi/Ti \\  system
\end{tabular}
&
$\begin{pmatrix}\pm1
&0\\0&0\end{pmatrix}$
&
\small{$\left\{
\begin{pmatrix}
\pm1
&b\\
0&a
\end{pmatrix}
\ \middle|\
a\neq 0
\right\}$}
&
\begin{tabular}{c}
$\Omega_1=\Omega_2=0$;\\
$\Omega_2=0,\quad|\Omega_1|=r>0$;\\
$\Omega_2\neq0$
\end{tabular}
\\
\hline
\begin{tabular}{c}
Compactified\\
Brinkmann\\
system
\end{tabular}
&
$\begin{pmatrix}0&0\\0&0\end{pmatrix}$
&
$GL(2,\mathbb R)$
&
\begin{tabular}{c}
$\Omega_1=\Omega_2=0$;\\
$(\Omega_1,\Omega_2)\neq(0,0)$
\end{tabular}
\\
\hline
\end{tabular}
\caption{Four inequivalent cases of $2\times 2$ matrix on the right-hand-side of the second line in \eqref{two-scales} }
\label{tab:stabilizer-orbits}
\end{table}

If restrictions are imposed on the allowed values of $\Omega_i$, for instance $\Omega_i\geq 0$, some of these transformations cease to be symmetries of the two-scale system. However, no such restriction will be imposed in the present section.

%%%%%%%%%%%
\subsection{The canonical Killing vector field and its orthogonal foliation}
%%%%%%%%%%%
\begin{defi}
Let $G$ be a conformal metric and $\Omega_1,\Omega_2$ two Weyl densities of weight one.
\begin{itemize}
    \item 
The vector field $V$ defined as 
\begin{align}\label{defV}
   V=\Omega_1\, \mathrm{grad}_G(\Omega_2)-\Omega_2\,\mathrm{grad}_G(\Omega_1)\,,
\end{align}
 is Weyl-invariant and will be called the canonical vector field, where $\mathrm{grad}_G(f)$ is the vector field such that $G\big(\mathrm{grad}_G(f),\,\cdot\,\big)=df$.
\item Under the regularity assumptions \eqref{regularity}, the family of hypersurfaces 
\begin{align}\label{projective}
    a\,\Omega_1+b\,\Omega_2=0\,,
\end{align}
where $[a:b]$ are homogeneous coordinates on $\mathbb{RP}^1$, defines a singular foliation of codimension one, regular on $M\setminus\mathcal{C}$, which we call the canonical singular foliation. When restricted to $M\setminus\mathcal{C}$, we simply refer to it as the canonical foliation.
\end{itemize}
\end{defi}

We will encounter singular foliations throughout the paper, but let us stress that we use this term in the elementary sense of a family of hypersurfaces sharing a common singular locus and defining a regular codimension-one foliation away from it.  Moreover, the regularity assumptions \eqref{regularity} ensure that the singular locus is a smooth submanifold. Thus, the singularities of the foliations considered in this paper are extremely mild.
Let us also note that, on $M\setminus\mathcal{C}$, the canonical vector field $V$ is orthogonal to the leaves of this foliation, since its dual one-form $V^\flat=\Omega_1d\Omega_2-\Omega_2d\Omega_1$ annihilates their tangent vectors.

We now discuss the conditions that the two-scale system \eqref{two-scales} imposes on the canonical vector field and the canonical singular foliation.
\begin{prop}\label{prop-conf-killing}
Let $(G,\Omega_1,\Omega_2)$ be a solution of the two-scale system. Then
\begin{enumerate}
    \item The canonical vector field $V$ is a conformal Killing vector field. More precisely, it satisfies
\begin{align}\label{prop-conf-killing-eq}
\mathcal{L}_V G=2\,(\Omega_2 \rho_1-\Omega_1\rho_2)G\,.
\end{align}
\item  On the domain where $V^2:=G(V,V)\neq0$, the leaves of the canonical foliation are totally umbilic, i.e. the trace-free part of their second fundamental form vanishes.
\end{enumerate}
\end{prop}

\begin{proof}
1) The definition \eqref{defV} of the canonical vector field $V$ together with the first line in \eqref{two-scales} give 
\begin{align}
    \nabla_{A}V_B+\nabla_BV_A=2(\Omega_1\nabla_A\nabla_B\Omega_2-\Omega_2\nabla_A\nabla_B\Omega_1)=-2(\Omega_1\rho_2-\Omega_2\rho_1)G_{AB}\,.
\end{align}
2) Taking the covariant Hessian of $a\,\Omega_1+b\,\Omega_2$ and using the two-scale equations, we obtain
\begin{align}
    a\,\nabla_A\nabla_B\Omega_1+b\,\nabla_A\nabla_B\Omega_2=-(a\,\Omega_1+b\,\Omega_2)P_{AB}-(a\rho_1+b\rho_2)G_{AB}\,.
\end{align}
The term proportional to the Schouten tensor vanishes on the hypersurface \eqref{projective}. The tangential restriction of this Hessian is therefore proportional to the induced metric. Since the second fundamental form is proportional to this restriction, the hypersurface is totally umbilic.
\end{proof}
In the context of conformal geometry, the fact that multiple almost-Einstein scales give rise to conformal Killing vector fields was observed in \cite{gover2010class}. The same reference also contains an observation that leads to the following equivalent definition of two-scale systems. 
\begin{prop}\label{firstrem}
    The two-scale system \eqref{two-scales} admits an equivalent reformulation, 
    \begin{align}\label{two-scales-equiv}
    \begin{split}
        &\nabla_A \nabla_B \Omega_1 + P_{AB}\Omega_1 + G_{AB}\rho_1 = 0\,,\\
        & \left(\nabla_{A}V_B+\nabla_BV_A\right)\big|_{\mathrm{tf}}=0\,,\\
        &\Omega_i\rho_j+\Omega_j\rho_i+G^{AB}\partial_A\Omega_i\,\partial_B\Omega_j=\lambda_{ij}\,,\qquad i,j=1,2\,,
    \end{split}
    \end{align}
   where $|_{\mathrm{tf}}$ denotes taking the trace-free part and $V$ is the canonical vector field given by~\eqref{defV}.
\end{prop}
In other words, in the definition of a two-scale system the almost-Einstein equation for one of the two scales can be replaced with the conformal Killing equation for the canonical vector field $V$. 

\begin{proof}By using the definition of $V$ and the almost-Einstein equation for $\Omega_1$, one finds
\begin{align}
    \begin{split}\nabla_{A}V_B+\nabla_BV_A&=2(\Omega_1\nabla_A\nabla_B\Omega_2-\Omega_2\nabla_A\nabla_B\Omega_1)
    =2\Omega_1(\nabla_A\nabla_B\Omega_2+\Omega_2P_{AB})+2\Omega_2\rho_1 G_{AB}\,.
    \end{split}
\end{align}
The conformal Killing equation for $V$ implies that 
    \begin{align}
        \Omega_1(\nabla_{A}\nabla_B\Omega_2+P_{AB}\Omega_2)|_{\mathrm{tf}}=0\,.
    \end{align}
This proves the equivalence in the region where $\Omega_1\neq0$. To extend this relation to $\Omega_1=0$, we differentiate the above equation covariantly and then restrict to the hypersurface $\Omega_1=0$. Using the regularity assumption \eqref{regularity}, we recover the desired relation on the hypersurface.
\end{proof}

Note that the squared norm of the canonical vector field is a weight-two Weyl density which can be represented as the following quadratic polynomial in the two scales
\begin{align}\label{V2}
V^2=\lambda_{11}(\Omega_2)^2+\lambda_{22}(\Omega_1)^2-2\lambda_{12}\Omega_1\Omega_2\,,
\end{align}
provided $(G,\Omega_1,\Omega_2)$ is a solution of the two-scale system. As one can see, the norm of the canonical vector field vanishes:
\begin{itemize}
    \item on four
    leaves when $\det(\lambda)<0$ (compactified Fefferman-Graham system),
    \item on two leaves when $\det(\lambda)=0$ but $\lambda\neq 0$ (compactified Spi/Ti ambient system),
    \item only on $\mathcal C$ when $\det(\lambda)>0$ (compactified Einstein degression),
    \item everywhere when $\lambda=0$ (compactified Brinkmann system).
\end{itemize}
Away from $\mathcal C$, the condition $V^2=0$ on a leaf of the canonical foliation is equivalent to that leaf being a null hypersurface, i.e. the pullback of the conformal metric $G$ to it is degenerate, with the restriction of $V$ to the leaf spanning its kernel.

%%%%%%%%%%%
\subsection{\texorpdfstring{$\Omega_1\neq0$}{Omega non 0} domain}\label{sec:omeganon0}
%%%%%%%%%%%

\begin{defi}Let $G$ be a conformal metric and $\Omega_1,\Omega_2$ be two Weyl densities of weight one. On the domain $\Omega_1\neq0$ of $M$, one introduces the Weyl-invariant objects
    \begin{align}\label{int-projection1}
        \tilde G_{AB}:=\Omega_1^{-2}G_{AB},\qquad u:= \Omega_1^{-1}\Omega_2\,,
    \end{align}
    which will be called, respectively, the ambient metric and the level function. 
\end{defi}
The term ``level function'' is justified by the fact that the level hypersurfaces $u=\mathrm{const}$ are precisely the leaves of the canonical foliation; see \eqref{projective}.
\begin{prop}\label{prop-omega1>0}
The two scale system~\eqref{two-scales} subject to the additional condition $\Omega_1 \neq 0$ is equivalent to the following system 
    \begin{align}\label{gauge-fixed1}
        \begin{split}
        &\tilde P_{AB}=-\tfrac12\lambda_{11}\tilde G_{AB},\\
        &\tilde \nabla_A\tilde \nabla_Bu=(\lambda_{11}u-\lambda_{12})\tilde G_{AB},\\
        &\tilde G^{AB}\partial_A u\,\partial_B u=P_2(u)
        \end{split}
    \end{align}
for the level function $u$ and the ambient metric $\tilde G$, where $P_2$ is a polynomial of degree at most two given by
    \begin{align}\label{P2pol}
        P_2(u):=\lambda_{11}u^2-2\lambda_{12}u+\lambda_{22}\,.
    \end{align}
In this formulation, the canonical vector field $V$ becomes: 
\begin{itemize}
    \item the gradient of the level function (i.e. $V=\mathrm{grad}_{\tilde G}(u)$ or, equivalently, $V^\flat=du$),
    \item a conformal Killing vector field of the ambient metric, $\mathcal{L}_V\tilde G=2(\lambda_{11}u-\lambda_{12})\tilde G$. 
\end{itemize}
Its norm square $V^2$ is equal to the polynomial $P_2(u)$ of the level function and is therefore constant on each leaf $u=\mathrm{const}$ of the canonical foliation.

\end{prop}
In what follows, we call the system of equations \eqref{gauge-fixed1} for $\tilde G$ and $u$ the \textit{ambient system}. Let us note that the ambient system, with the tildes removed and $u$ replaced by $\Omega_2$, can be obtained by fixing the Weyl gauge symmetry of the two-scale system with the gauge-fixing condition $\Omega_1=1$. 
Indeed, in this gauge, the second line of $\eqref{two-scales}$ immediately yields  $\rho_1=\frac{1}{2}\lambda_{11} $, $\rho_2=\lambda_{12}-\frac{1}{2}u\lambda_{11}$. Note that this gauge $\Omega_1=1$ is reachable. The system \eqref{gauge-fixed1} will be analysed in the following sections, case by case, for different choices of the matrix $\lambda$. But note already that the first equation in \eqref{gauge-fixed1} asserts that $\tilde G$ is an Einstein metric.

Let us comment on a geometrical interpretation of the last equation in \eqref{gauge-fixed1}. The regularity assumptions \eqref{regularity} imply that $du\neq0$. Hence, the map $\pi:M\to \fR$ determined by $\pi^* w=u$, where $w$ is the standard coordinate on the real line $\fR$, defines a fibre bundle at least locally. Of course, its fibres are precisely the leaves of the canonical foliation. The polynomial $P_2(u)$ seen as a function on $M$ is the pullback of $P_2(w)$ defined on the base $\fR$. In fact $P_2(w)$ is naturally interpreted as the only component of the inverse metric $P=P_2(w)\,\dl{w}\tensor \dl{w}$. From this perspective, the last equation in \eqref{gauge-fixed1} tells us that the inverse of $\tilde G$ is a projectable cometric on $M$ whose projection is precisely $P$. Moreover, it follows from $\tilde G(V,\cdot)=du$ that $V$ is a projectable vector field whose projection is $P_2(w)\dl{w}$.
\\

Now let us introduce a useful coordinate system adapted to the ambient system \eqref{gauge-fixed1}. Here and in what follows, indices denoted by lowercase Latin letters from the middle of the alphabet (such as $m,n,\ldots$) will take $d$ values.
\begin{prop}\label{prop-local-solution0}
Let $(\tilde G,u)$ be a solution of the system \eqref{gauge-fixed1}. Then any point has an open neighbourhood and a coordinate system  $(\rho,v,x^m)$ in which it takes the form
\begin{align}\label{catalano-coord}
        \begin{split}
         \tilde G&=2\,\partial_\rho u\, dv\,d\rho\,+\,P_2(u)\,dv^2\,+\,\partial_\rho u\,\,h_{mn}(\rho,x)\,dx^mdx^n\,,\\
         &u=u(\rho,v)\,,\quad \partial_vu(\rho,v)=P_2(u)\,,\quad u(\rho,0)=\rho\,,
         \end{split}
\end{align}
     where $\det h_{mn}\neq0$.
\end{prop}

The proof is given in Appendix~\bref{proofPropproplocal0}. Note that, interpreting \eqref{catalano-coord} as an ansatz for a solution of this system \eqref{gauge-fixed1}, the second and third equations of the system are trivially satisfied, while the first imposes non-trivial constraints on $h_{mn}$.
\\

By definition, a homothetic vector field is a conformal Killing vector field such that the conformal factor defined by the corresponding conformal isometry is constant. In particular, a metric is said self-similar if the Lie derivative of the metric is equal to twice the metric. Homothetic vector fields play a fundamental role both in conformal geometry and in general relativity. In mathematical relativity, they often arise in the context of self-similar limits of families of Einstein metrics \cite{Rodnianski:2017cvt}. Furthermore, they are central to the analysis of critical phenomena in gravitational collapse \cite{Gundlach:1999cu}. In conformal geometry, the ambient metric in the Fefferman-Graham construction is both Ricci-flat and self-similar \cite{Fefferman-Graham:1985ambient,FeffermanGraham2012}. As a side remark, let us observe that Ricci flatness and homothety symmetry of the ambient metric necessarily come simultaneously for two-scale systems. 
\begin{rem}\label{sideremark}
Under the additional restriction $\Omega_1\neq0$, the following conditions on the two-scale system \eqref{two-scales} are equivalent:
\begin{enumerate}
    \item The top diagonal entry of the constant matrix $\lambda$ vanishes ($\lambda_{11}=0$),
    \item The ambient metric $\tilde G:=\Omega_1^{-2}G$ is Ricci-flat ($\tilde R_{AB}=0$),
    \item The  canonical  vector field $V$ is a homothetic vector field of the ambient metric ($\mathcal{L}_V\tilde G=2\tilde\omega\tilde G$ with $d\tilde\omega=0$).
\end{enumerate}
\end{rem}

%%%%%%%%%%%
\subsection{Induced metrics on the leaves of the canonical foliation}
%%%%%%%%%%%
In  Proposition \bref{prop-local-solution0}, we have already introduced a coordinate system adapted to the ambient system \eqref{gauge-fixed1}. Now we show that, under an additional condition, we can introduce a different and somewhat more useful coordinate system adapted to the canonical foliation. Here and in what follows, the indices denoted by lowercase Latin letters at the beginning of the alphabet ($a,b,\ldots$) will take $d+1$ values. 
\begin{prop}\label{prop-local-solution} Let  $(\tilde G,u)$ be a solution of the system~\eqref{gauge-fixed1}.
Any point where $P_2(u)\neq0$ has an open neighbourhood and a coordinate system $(u,y^a)$ therein such that the ambient metric takes the following form:
\begin{align}\label{generic-tilde-g}
    \tilde G=\frac{du^2}{P_2(u)}
+\big|P_2(u)\big|\,\gamma_{ab}(y)\,dy^a\,dy^b\,,
\end{align}
where $\gamma$ is a $(d+1)$-dimensional Einstein metric whose Ricci tensor satisfies
\begin{align}\label{two-scales-einstein}
        R_{ab}[\gamma]\,=\,-\,d\,\sgn(P_2)\,\det\lambda\,\gamma_{ab}\,,
    \end{align}
and the signature of $\gamma$ is  
$(p,q+1)$ when $P_2(u)>0$ and $(p+1,q)$ when $P_2(u)<0$. Moreover, in this local coordinate system, the canonical vector field reads as
\begin{equation}\label{canvectf}
    V=P_2(u)\frac{\partial}{\partial u}\,.
\end{equation}
\end{prop}
Note that our conventions for the cosmological constant $\Lambda$ are $R_{ab}-\frac{1}{2}R\gamma_{ab}-\Lambda
\gamma_{ab}=0$ and hence \eqref{two-scales-einstein} gives
\begin{align}\label{cosm-const}
    \Lambda
    =\sgn(P_2)\,\frac{d(d-1)}{2}\det\lambda\,.
\end{align}
The idea of the proof is to show that under the stated assumptions, coordinates $u,y^a$ can be chosen in such a way that $\tilde{G}_{ua}=0$. Then equations \eqref{gauge-fixed1}
imply~\eqref{generic-tilde-g}. Details are given in Appendix~\bref{proofPropproplocal}. 
Metrics of the form \eqref{generic-tilde-g} were already studied by Brinkmann \cite{Brinkmann:1925fr}, who showed that they locally characterize Einstein metrics admitting a proper conformal map to another Einstein metric, where proper means that the gradient of the conformal factor has nonzero squared norm. In the next section, we explain how this property is related to the field redefinitions of the two-scale system introduced in Section~\bref{sec:symmetries}.

The metric $\gamma=\gamma_{ab}(y)\,dy^a\,dy^b$ coincides, up to a positive factor, with the metric induced on the leaf $u=\mathrm{const}$. Therefore, with a slight abuse of terminology $\gamma$ will sometimes be called the \textit{induced metric}. Note that if $d+2\leq4$ and $P_2(u)\neq0$, then any solution of the system \eqref{gauge-fixed1} is locally equivalent to a maximally symmetric spacetime, since condition \eqref{two-scales-einstein} uniquely fixes the metric $\gamma$ locally.

Finally, we note that \eqref{generic-tilde-g} with $\gamma$ an Einstein metric satisfying \eqref{two-scales-einstein}, can be interpreted as an ansatz for a solution of the ambient system \eqref{gauge-fixed1}. All the equations of the system are trivially satisfied, in contrast to the coordinate system in \eqref{catalano-coord}, where the Einstein condition on the ambient metric imposed additional constraints. Thus, given the Einstein metric $\gamma$, the ambient system provides, locally and under the assumption $P_2(u)\neq0$, an ambient space with Einstein metric $\tilde G$. This space is foliated by the level hypersurfaces $u=\mathrm{const}$ endowed with Einstein metrics proportional to $\gamma$.
%%%%%%%%%%%
\subsection{Fractional linear transformations and equivalent formulations}\label{sec:equiv}
%%%%%%%%%%%

We now analyze how the transformation \eqref{GLtransfos} described in Section~\bref{sec:symmetries} is expressed in terms of the system~\eqref{gauge-fixed1}. Because \eqref{GLtransfos} commutes with Weyl transformations, its action on level function $u$ and ambient metric $\tilde{G}$ is well-defined
provided it preserves the condition $\Omega_1\neq0$. More precisely, the action can be written as follows
\begin{align}\label{sym-int1}
    u\to u'=\frac{A_{2}{}^{1}+A_{2}{}^{2}u}{A_{1}{}^{1}+A_{1}{}^{2}u},\qquad \tilde{G}\to\tilde G'=(A_{1}{}^{1}+A_{1}{}^{2}u)^{-2}\tilde G\,,
\end{align}
where 
\begin{align}
    A=\begin{pmatrix}
A_1{}^{1}&A_{1}{}^{2}\\
A_{2}{}^{1}&A_{2}{}^{2}
\end{pmatrix}\in GL(2,\mathbb{R})
\end{align}
and preserving the condition $\Omega_1\neq0$ gives $A_{1}{}^{1}+A_{1}{}^{2}u\neq0$ which is necessary for writing down \eqref{sym-int1}.

The level function $u$ is nothing but the inhomogeneous coordinate $\Omega_2/\Omega_1$ in a neighborhood $\Omega_1\neq 0$ on the projective line $\mathbb{RP}^1$ which is the projectivisation of the $(\Omega_1,\Omega_2)$-plane formed by the two scales.\footnote{Note that the above action \eqref{sym-int1} of $GL(2,\mathbb{R})$ is not effective, since $A$ and $-A$ produce the same transformation. However, it descends to an effective action of  the quotient group $GL(2,\mathbb{R})/\mathbb{Z}_2$ by the central subgroup $\mathbb{Z}_2=\{I,-I\}$. Furthermore, the latter admits the following decomposition $GL(2,\mathbb{R})/\mathbb{Z}_2\cong \mathbb{R}_{>0}\times PGL(2,\mathbb{R})$, where only the M\"obius group $PGL(2,\mathbb{R})$ acts effectively on $u$ while $\mathbb{R}_{>0}$ only acts on the ambient metric by a mere rescaling.} The polynomial \eqref{V2} can be thought as the projectivisation of $P_2(u)$.
If $A\in \mathrm{Stab}(\lambda)$, the transformation \eqref{sym-int1} is a symmetry of the system of equations~\eqref{gauge-fixed1}, i.e. such a transformation maps solutions to solutions. A natural question then arises: can this transformation be related to a symmetry of a given solution?

\begin{prop}\label{prop:diffeo}
Let $(\tilde G_{AB},u)$ be a solution of the ambient system \eqref{gauge-fixed1}. On an open set where $P_2(u)\neq0$, choose the coordinates $(u,y^a)$ of Proposition~\bref{prop-local-solution}. For $A\in GL(2,\mathbb R)$ the local diffeomorphism $\varphi$ of $M$ defined by
    \begin{align}
       \varphi^{*}u=\frac{A_{2}{}^{1}+A_{2}{}^{2}u}{A_{1}{}^{1}+A_{1}{}^{2}u}\,,\qquad \varphi^{*}y&^a=y^a 
    \end{align}
    is a conformal isometry of the ambient metric such that 
    \begin{align}
        \varphi^{*}\tilde G_{AB}=(A_{1}{}^{1}+A_{1}{}^{2}u)^{-2}\tilde G_{AB}\,.
    \end{align}
    if and only if $A\in\mathrm{Stab}(\lambda)$ and $\det(A)=\pm1$.
\end{prop} 
\noindent This proposition can be checked by inserting the coordinate transformation into the explicit form of the ambient metric \eqref{generic-tilde-g}. Note that this  argument applies only in the domain $P_2(u)\neq0$ and is therefore inapplicable to the case $\lambda=0$, for which $P_2(u)\equiv0.$

For $A\notin \mathrm{Stab}(\lambda)$, the transformation \eqref{sym-int1} produces a system of equations for $(\tilde G',u')$ equivalent to \eqref{gauge-fixed1}, but with coefficients $\lambda'=A\lambda A^T$ instead of $\lambda$. In particular, this transformation maps Einstein metrics to Einstein metrics, recovering Brinkmann’s conformal rescalings~\cite{Brinkmann:1925fr}. Although the general linear transformations look almost trivial for the full two-scale system, their realisation for the ambient system in the interior \eqref{gauge-fixed1} is not so obvious and can lead to equivalent systems that look superficially very different. 
As an illustration, in Section~\bref{sec: ads-ads} we demonstrate that the Fefferman--Graham ambient construction can be equivalently reformulated as an embedding of Einstein spaces not into a Ricci-flat, but into an Einstein ambient space.

%%%%%%%%%%%
\subsection{\texorpdfstring{$\Omega_1=\Omega_2=0$ submanifold}{Omega1=Omega2=0 submanifold}}\label{app: coord-corner}
%%%%%%%%%%%
In this section, we set aside the condition $\Omega_1\neq0$ and analyze solutions of the two-scale system in a neighbourhood of the codimension-two submanifold \eqref{corner}, i.e. $\mathcal{C}=\{\Omega_1=\Omega_2=0\}$.
If $(G,\Omega_1,\Omega_2)$ is a given solution of the two-scale system \eqref{two-scales}, then $\mathcal C$ is naturally endowed with the induced conformal metric given by the pullback of $G$. Unlike in regions where one of the scales is nonzero, neither scale can be used to fix the Weyl freedom on $\mathcal C$. Thus, the geometry induced on $\mathcal C$ is genuinely conformal.  Suppose $(n_+,n_-,n_0)$ is the index of inertia for $\lambda$ (i.e. the number of positive, negative, and zero eigenvalues, respectively). Then the pullback of the conformal metric $G_{AB}$ on $\mathcal C$ has $n_0$ degenerate directions and, in the non-degenerate sector, has signature $(p+1-n_+-n_0,q+1-n_--n_0)$, as will become clear below.

\begin{lemma}\label{lemma-rho}
Let $(G,\Omega_1,\Omega_2)$ be a given solution of the two-scale system. Then, in a neighbourhood of any point from $\mathcal{C}$, there exists a function $\alpha(x)$ such that the Weyl-transformed fields $G'=e^{2\alpha(x)}G$ and $\Omega_i'=e^{\alpha(x)}\Omega_i$ satisfy $\rho'_i=0$, i.e. 
\begin{align}
    \nabla^{'2}\Omega'_i+P'\,\Omega'_i=0\,,\qquad i=1,2\,,
\end{align}
and, moreover,
\begin{align}
[\,\mathrm{grad}_{G'}\,\Omega_1',\mathrm{grad}_{G'}\,\Omega_2'\,]=0\,.
\end{align}
\end{lemma}
\noindent The proof of this lemma can be found in Appendix~\bref{app: rho-gauge}.\\
    
 It follows from~\eqref{two-scales} that  the Weyl-transformed fields of the above lemma satisfy
\begin{align}\label{two-scales-without-rho}
    \begin{split}
        &\nabla_A \nabla_B \Omega_{i} + P_{AB}\Omega_i=0\,,\\
        &G^{AB}\partial_A\Omega_i\partial_B\Omega_j=\lambda_{ij}\,,
    \end{split}
\end{align}
where here and henceforth we omit the primes for brevity. From a field-theoretical perspective, this means that there exists a local Weyl gauge in which the two-scale system takes the form \eqref{two-scales-without-rho}.
Note also that, in this gauge, the canonical vector field $V$ is a Killing vector field, as follows from \eqref{prop-conf-killing-eq}. We use this fact together with the commutativity of the gradient vector fields 
to construct a convenient coordinate system near $\mathcal{C}$:
\begin{prop}\label{prop:corner-coord}
Let $(G,\Omega_1,\Omega_2)$ be a solution of the two-scale system~\eqref{two-scales}, and let $p$ be a point of the submanifold $\mathcal C$. Assume that the solution is written in the Weyl representative provided by Lemma~\bref{lemma-rho} near $p$. Then there exist local coordinates $(t^i,x^m)$ near $p$ in which the solution takes the form
\begin{equation}\label{G-near-corner}
\begin{gathered}
\Omega_i=\lambda_{ij}t^j+g_i(x)\,,\\
G=\lambda_{ij}dt^idt^j+2dt^idg_i(x)+h_{mn}(\lambda_{ij}t^it^j+2g_i(x)t^i,x)dx^mdx^n+\mathcal{R}\,, 
\end{gathered}
\end{equation}
where $g_i(x)$ are smooth functions, $h_{mn}(s,x)$ is a smooth one-parameter family of $d$-dimensional symmetric rank-2 tensors, and the smooth remainder  $\mathcal{R}=\mathcal{R}_{mn}(t,x)dx^mdx^n$ is invariant under $V$ and vanishes faster than any power of $|\Omega_1|+|\Omega_2|$ as one approaches $\mathcal{C}$.
\end{prop}
\noindent Notice that the remainder $\mathcal R$ is absent when working with formal expansions along $\mathcal C$ and vanishes identically for analytic solutions.

\begin{proof}
 Since the gradient vector fields of $\Omega_i$ commute, we can choose a coordinate system $(t^i,x^m)$ such that $\mathrm{grad}_G\,\Omega_i=\frac{\partial}{\partial t^i}$ 
    The second equation in \eqref{two-scales-without-rho} then yields
    \begin{align}
        \frac{\partial\Omega_i}{\partial t^j}=\lambda_{ij}\qquad \Rightarrow\qquad \Omega_i=\lambda_{ij}t^j+g_i(x)\,.
    \end{align}
    Substituting it back to the second equation in \eqref{two-scales-without-rho}, we obtain a metric in the form.
    \begin{align}
    G=\lambda_{ij}dt^idt^j+2dt^idg_i(x)+ h_{mn}(t,x)dx^mdx^n\,. 
    \end{align}
The condition that $V$ be a Killing vector field imposes the following equations on $ h_{mn}$
\begin{align}\label{Vh}
    \left(\Omega_1\frac{\partial}{\partial t^2}-\Omega_2\frac{\partial}{\partial t^1}\right) h_{mn}(t,x)=0\,.
\end{align}
Since the function $s:=\lambda_{ij}t^it^j+2g_i(x)t^i$ is preserved by the vector field $V$,  away from $\mathcal C$ the method of characteristics gives solutions locally of the form $h_{mn}\big(s(t,x),x\big)$. Different connected components of a level set of $(s,x^m)$, i.e. a locus such that $(s,x^m)=(s_0,x^m_0)$, may carry different such functions. Smoothness across $\mathcal C$ requires these branches, expressed in the regular coordinates $(t^i,x^m)$, to match to every order. Thus any branch-dependent contribution must vanish to infinite order along $\mathcal C$.
\end{proof}

\begin{cor}\label{cor:appA}
    Suppose $\det\lambda\neq0$. Then
    \begin{align}
        \Omega_i=\lambda_{ij}t^j+g_i(x)
    \end{align}
    defines a smooth change of coordinates from $(t^i,x^m)$ to $(\Omega_i,x^m)$.  Consequently, after redefining $h_{mn}$, the metric $G$ in \eqref{G-near-corner} takes the form
        \begin{align}\label{lambdaomegasquared}
        G=\lambda^{ij}d\Omega_id\Omega_j+h_{mn}(\lambda^{ij}\Omega_i\Omega_j,x)dx^mdx^n+\mathcal{R}\,,
    \end{align}
    where $\lambda^{ij}\lambda_{jk}=\delta^{i}_k$.
\end{cor}

Let us finally comment on the geometrical interpretation of the second equation in~\eqref{two-scales-without-rho}. If we restrict ourselves to a neighbourhood of $\mathcal{C}$ where $d\Omega_1\wedge d\Omega_2\neq 0$ then the functions $\Omega_i$ determine a map $M \to \fR^2$ which defines a fibre bundle, at least locally. Then the second equation from~\eqref{two-scales-without-rho} means that the cometric $G^{AB}$ is projectable and its projection is the twice-contravariant symmetric tensor $\lambda_{ij}\dl{\Omega_i}\tensor \dl{\Omega_j}$ defined on $\fR^2$. If $\det \lambda \neq 0$ this tensor can be regarded as a cometric on $\fR^2$. It is also easy to see that canonical vector field $V$ is projectable. Because $G^{AB}$ is projectable the projected $V$ is determined by the same equation
\eqref{defV} but now on $\fR^2$ with the cometric $\lambda$ and coordinate functions $\Omega_i$.

%%%%%%%%%%%%%%%%%%%%%%%%%%%%%%%%%%%%%%%%%%%%%%%%%%%%
\section{Compactified Fefferman-Graham system}\label{FGsection}
%%%%%%%%%%%%%%%%%%%%%%%%%%%%%%%%%%%%%%%%%%%%%%%%%%%%
The properties of the two-scale system~\eqref{two-scales} depend crucially on the congruence orbit of $\lambda_{ij}$ and hence it is natural to study the systems corresponding to different orbits separately. In this section we concentrate on the orbit containing 
\begin{align}\label{compFGlambda}
    \lambda=
    \begin{pmatrix}
        0 & -1\\
        -1 & 0
    \end{pmatrix}\,.
\end{align}
\begin{defi}\label{def-comp-fg}
We call the two-scale system \eqref{two-scales} with $\lambda_{ij}$ given by \eqref{compFGlambda} and with the Weyl densities subject to
\begin{align}
    \Omega_i\geq0,\qquad i=1,2\,,
\end{align}
the compactified Fefferman--Graham system.
\end{defi}
In studying this system we assume 
$M$ to be   a manifold with corners, with $\Omega_i$ serving as defining functions for the two boundary faces and with the corner $\mathcal C$ as their intersection. The regularity assumptions \eqref{regularity} ensure that the faces are smooth and intersect transversally along the corner $\mathcal C$. Away from $\mathcal C$, both faces are null hypersurfaces to which the canonical conformal Killing vector field $V$ is normal and tangent; hence they are conformal Killing horizons. The corner $\mathcal C$ is a conformal manifold of signature $(p,q)$. The choice of $\lambda_{12}$ merely fixes the cosmological constant of the metric $\gamma$ induced on the leaves of the canonical foliation, see \eqref{cosm-const}, and so this choice does not play an essential role.
\begin{rem}
    If we drop the restrictions $\Omega_i\geq0$ and let $(G,\Omega_1,\Omega_2)$ be a solution that  extends smoothly across the hypersurfaces $\Omega_i=0$, then $\mathcal C$ is the bifurcation surface of a bifurcate conformal Killing horizon, and the domain $\Omega_i>0$ is locally one of the four regions separated by these two null hypersurfaces. In this section, we impose $\Omega_i\geq0$ to select one of these regions, the others admitting an analogous analysis.
\end{rem}

%%%%%%%%%%%
\subsection{\texorpdfstring{$\Omega_1>0$}{Omega>0} domain}
%%%%%%%%%%%
To interpret the system introduced in Definition~\bref{def-comp-fg} and explain its name, it is convenient, as in Section~\bref{sec:omeganon0}, to impose the additional condition $\Omega_1>0$ and rewrite the system in terms of the ambient metric $\tilde G$ and the level function $u$, which, as a consequence of $\Omega_2\geq0$, is restricted to $u\geq0$. 
Substituting \eqref{compFGlambda} into \eqref{gauge-fixed1}, we obtain the following equations:
\begin{align}\label{gauge-fixed}
        \begin{split}
            &\tilde P_{AB}=0,\\
            &\tilde\nabla_A\tilde\nabla_B u=\tilde G_{AB},\\
            &\tfrac{1}{2}\tilde G^{AB}\partial_Au\,\partial_Bu=u\,.
        \end{split}
\end{align}
Note that the second condition is equivalent to the homothety condition
\begin{align}
    \mathcal{L}_V\tilde G_{AB}=2\tilde G_{AB}
\end{align}
for the canonical vector field $V=\mathrm{grad}_{\tilde G}(u)$. Furthermore, the third condition is equivalent to $u=V^2/2$. Thus, we recover the standard Fefferman--Graham system \cite{Fefferman-Graham:1985ambient, FeffermanGraham2012} (more precisely, a straight ambient metric) for $\Omega_1>0$; see e.g. the appendix of \cite{Bekaert:2017bpy} for a concise presentation of the Fefferman--Graham system from a similar perspective. Proposition \bref{prop-omega1>0}  then implies that this two-scale system is the compactified Fefferman-Graham system of Definition \bref{def-comp-fg} if and only if the ambient metric is both Ricci-flat and self-similar.

The compactified Fefferman--Graham system defined for $\Omega_1\geq0$ , 
may be viewed as a conformal compactification of the ambient system \eqref{gauge-fixed}. Recall that, following Penrose \cite{Penrose:1962ij} and setting aside global aspects, a conformal compactification of a pseudo-Riemannian manifold $\tilde M$ with metric $\tilde G$ is a manifold with boundary admitting a boundary defining function $\Omega$ such that the rescaled metric $G:=\Omega^2\tilde G$ extends smoothly and non-degenerately to the boundary. In our setting, let $(\tilde G,u)$ be a solution of the ambient system \eqref{gauge-fixed} defined on a manifold $\tilde M$ and such that $u\geq 0$. Suppose that $\tilde M$ can be identified with the interior of a manifold with boundary $M$, i.e. $\tilde M=Int(M)$, equipped with a boundary defining function $\Omega_1$ such that $G:=\Omega_1^2\tilde G$ and $\Omega_2:=\Omega_1u$ extend smoothly to the boundary, with the extension of $G$ being nondegenerate and with $\Omega_1$ and the extension of $\Omega_2$ satisfying the regularity assumptions~\eqref{regularity}. It follows that $(G,\Omega_1,\Omega_2)$ defines a solution of the compactified Fefferman--Graham system on $M$. Conversely, any solution of the compactified Fefferman--Graham system on a manifold with boundary $M$, with $\Omega_1$ being a boundary defining function, can be restricted to $\Omega_1>0$ to give a solution $(\tilde G,u)=(\Omega_1^{-2}G,\Omega_1^{-1}\Omega_2)$ of the ambient system. Thus, solutions of the compactified Fefferman--Graham system correspond to solutions of the ambient system admitting such a conformal compactification. The above correspondence can be promoted from the level of solutions to that of equations by employing the notion of compactification of a gauge system introduced in~\cite{Grigoriev:2025gvk,Dneprov:2026muy}.

    \begin{rem}
  In a recent work \cite{Mars:2025lgr}, a version of the conformally compactified Fefferman--Graham system, closely related to ours, has been proposed. There, the authors consider a triple $(G,\Omega,X)$, where the scale $\Omega$ satisfies the almost-Einstein equation with respect to the conformal metric $G$ while $X$ is a conformal Killing vector field admitting a bifurcation surface and satisfying
$\mathcal{L}_X\Omega=\big(\frac{1}{d+2}\mathrm{div}(X)-1\big)\Omega$.
It appears that this and our compactified Fefferman-Graham system are locally equivalent under suitable regularity assumptions, see  Propositions~\bref{prop-conf-killing} and \bref{firstrem}. In particular, one can check that the canonical vector field $V$ from Proposition~\bref{prop-conf-killing} satisfies $\mathcal{L}_V\Omega_1=\big(\frac{1}{d+2}\mathrm{div}(V)-1\big)\Omega_1$. Moreover, the corner $\Omega_1=\Omega_2=0$ is the bifurcation surface for the  canonical vector field $V$. Let us stress, however, that the above definition of the compactified Fefferman-Graham system has a manifest  $\mathbb{Z}_2$-symmetry $\Omega_1\leftrightarrow\Omega_2$, to be discussed in Section~\bref{sec:Z2},and is a particular case of the more general two-scale system where $\lambda_{ij}$ is not necessarily of the form~\eqref{compFGlambda}.
\end{rem}

\begin{example}
Suppose the domain $\Omega_1>0$ is isometric to Minkowski spacetime with metric $\tilde G_{AB}=\eta_{AB}$. The general solution of equations \eqref{gauge-fixed} is given by
\begin{align}
    u(x)=\frac{1}{2}\eta_{AB}(x^A+a^A)(x^B+a^B)
\end{align}
where $x^A$ are Cartesian coordinates and $a^A$ are the components of an arbitrary constant vector. In other words, there is a one-to-one correspondence between the zero loci of all possible solutions $u(x)$ of \eqref{gauge-fixed}, with $\tilde G_{AB}$ being the Minkowski metric, and the null cones in Minkowski spacetime; see, e.g., \cite{Bekaert:2026cvx} for a pedagogical exposition of this well-known correspondence from a different perspective.
\end{example}

By Proposition~\bref{prop-local-solution}, any given solution $(\tilde G,u)$ of the ambient system \eqref{gauge-fixed} can be written, in a neighbourhood of any point where $u>0$, in local coordinates $(u,y^a)$ as
\begin{align}\label{twoscales-prop-u>0}
    \tilde G=\frac{1}{2u}\,du^2+2u\,\gamma_{ab}(y)dy^ady^b\,,
\end{align}
where $\gamma$ is an arbitrary $(d+1)$-dimensional metric satisfying 
\begin{align}\label{fg-two-scales-einstein}
    R_{ab}[\gamma]=d\,\gamma_{ab}\,.
\end{align}
This is of course the standard fact on the Fefferman-Graham ambient construction that $u=const\neq 0$ submanifolds are Einstein spaces with non-vanishing cosmological constant.

The following proposition is the essence of the Fefferman-Graham ambient metric approach to the near boundary analysis of the asymptotically AdS spacetimes:

    \begin{prop}\label{two-scales-prop-solution}
Let  $N =\fR^{d}\times \fR_{>0}$ be a near boundary coordinate patch of the asymptotically AdS spacetime with the metric $\gamma$ subject to \eqref{fg-two-scales-einstein}.  Let also $(\tilde G,u)$ be the corresponding ambient solution \eqref{twoscales-prop-u>0} defined on $\mathbb{R}_{u>0}\times N$. Then this solution continuously extends to $\mathbb{R}_{u \geq 0}\times N$, i.e. to the boundary $u=0$. Moreover, this boundary is null.
    \end{prop}

\begin{proof}
Near its asymptotic boundary, an asymptotically AdS metric can locally be brought to the normal form \cite{FeffermanGraham2012}
        \begin{align}\label{aads-normal-form}
            \gamma=\frac{1}{\sigma^2}\big(-d\sigma^2+\frac{1}{2}h_{mn}(\sigma,x)dx^mdx^n\big)\,.
        \end{align}
        where we have introduced coordinates $(y^a)=(\sigma,x^m)$, $\sigma>0$. Substituting this into \eqref{twoscales-prop-u>0} and performing the transformation 
        \begin{align}\label{coord-t-rho}
            t=\sqrt{u}/\sigma\,,\qquad \rho=\sigma^2\,,
        \end{align}
        the ambient metric takes the standard Fefferman-Graham form
        \begin{align}\label{twoscales-prop-answer}
            \tilde G=2\rho dt^2+2tdtd\rho+t^2h_{mn}(\sqrt{\rho},x)dx^mdx^n
            \end{align}
            which is continuous at $\rho=0,$ $t>0$, while the level function is given by 
            \begin{align}
                u=\rho t^2\,.
            \end{align}
            Thus, this solution extends straightforwardly to the region $u=0$ (actually $\rho=0$ in these coordinates).
            Note that the resulting metric is not necessarily smooth, since $h$ depends on $\sqrt{\rho}$.
    \end{proof}
Note that the resulting extensions are smooth if $h_{mn}(\sigma,x)$ in \eqref{aads-normal-form} is a smooth even function of its first argument. In this case, the extension defines a solution of the ambient system \eqref{gauge-fixed} on a manifold with boundary.
\begin{rem}\label{rem:FG-reverse}
A similar statement can be obtained by employing the coordinate system $(\rho,v,x^m)$ introduced in Proposition~\bref{prop-local-solution0}.  Consider 
a coordinate patch where $u$ has non-trivial zero loci and assume that we are given a smooth solution to the compactified Fefferman-Graham system defined on this patch. Because these coordinates are regular anywhere, including a neighbourhood of $u=0$, the solution can be written in these coordinates. For $\lambda_{ij}$ given by \eqref{compFGlambda}, it takes the form   
        \begin{align}
                     \tilde G&=e^{2v}\big(2\,dvd\rho+2\rho\, dv^2+h_{mn}(\rho,x)\,dx^mdx^n\big)\,,\qquad u=e^{2v}\rho\,.
        \end{align}
After the change of variables $t=e^v$ the metric takes the form \eqref{twoscales-prop-answer}, except that $h_{mn}$ now depends on $\rho$ rather than on $\sqrt{\rho}$, which is, of course, compatible with the assumption that the solution is smooth at $u=0$, i.e. at $\rho=0$ in these coordinates.  Passing to the coordinates $\sigma,x^m$ defined by the inverse of~\eqref{coord-t-rho} one finds that the metric induced on $u=const$ takes the form 
\begin{align}
                    \gamma=\frac{1}{\sigma^2}\big(-d\sigma^2+\frac{1}{2}h_{mn}(\sigma^2,x)dx^mdx^n\big)\,.
\end{align}
This is the standard form of the asymptotically AdS metric
with $h_{mn}(\sigma^2,x)$ being an even function of $\sigma$. Note that if $d$ is odd, such metrics have vanishing subleading mode, or equivalently vanishing Neumann data; see, for example, \cite{deHaro:2000xn,Grigoriev:2025gvk}.
\end{rem}

    Finally, a natural question is whether the solution constructed in Proposition~\bref{two-scales-prop-solution} can be lifted to a solution of the compactified Fefferman-Graham system. The answer is affirmative:
    \begin{cor}\label{coroldoublescalecoords}
    In the setting of Proposition~\bref{two-scales-prop-solution} there exist coordinates $(\Omega_1,\Omega_2,x^m)$ on   $\mathbb{R}_{\Omega_1 > 0}\times \mathbb{R}_{\Omega_2 > 0}\times\mathbb{R}^d$ such that $u=\Omega_1^{-1}\Omega_2$ and that the rescaled metric $G=\Omega_1^2\tilde G$ extends continuously as a non-degenerate metric to $\mathbb{R}_{\Omega_1 \geq 0}\times \mathbb{R}_{\Omega_2 \geq 0}\times\mathbb{R}^d$, i.e. to both boundary faces $\Omega_i=0$ and their intersection $\mathcal{C}$.
    \end{cor}
\begin{proof}
Indeed, introduce coordinate functions
 $ \Omega_1=t^{-1}$,   $\Omega_2=\rho t$ where $\rho,t$
 are part of the Fefferman--Graham coordinate $(\rho, t, x^m)$ system introduced in the proof of Proposition~\bref{two-scales-prop-solution}. In terms of the new coordinate system
 $(\Omega_1,\Omega_2,x^m)$ metric \eqref{twoscales-prop-answer} takes the following form
 \begin{align}\label{fg-fully-comp}
    G
    =\Omega_1^2\widetilde G
    =-2\,d\Omega_1\,d\Omega_2
      +h_{mn}\left(\sqrt{\Omega_1\Omega_2},x\right)
       dx^m dx^n\,.
\end{align}
It is clear that it extends continuously as a non-degenerate metric to both boundary faces and their intersection $\mathcal C$.
\end{proof}

If $h(\sigma,x)$ is smooth and even in its first argument, the extension yields a smooth solution of the two-scale system throughout the neighbourhood of the corner, including both boundary faces. In particular, combining this with the results of \cite{Fefferman-Graham:1985ambient,FeffermanGraham2012}, we can associate to every odd-dimensional conformal manifold a formal solution of the compactified Fefferman-Graham system. Note that Corollary~\bref{coroldoublescalecoords} 
does not contradict Corollary~\bref{cor:appA}, since the assumptions are different: one considers continuous solutions in the former and smooth ones in the latter. In fact, the analysis of smooth solutions to the two-scale system with $\det\lambda\neq0$ led to a metric of the form \eqref{lambdaomegasquared}, with $h$ depending on the product $\Omega_1\Omega_2$ rather than on its square root as for the continuous solutions in \eqref{fg-fully-comp}.

%%%%%%%%%%%
\subsection{\texorpdfstring{$\mathbb{Z}_2$ symmetry}{Z2 symmetry }}\label{sec:Z2}
%%%%%%%%%%%

The $\mathrm{Stab}(\lambda)$-symmetry \eqref{sym-int1}  of the ambient system can sometimes be reinterpreted as a symmetry of solutions. For instance, taking 
\begin{align}
    A=  
    \begin{pmatrix}
        0 & 1\\
        1 & 0
    \end{pmatrix}\in\mathrm{Stab}(\lambda),
\end{align}
and applying the general formula \eqref{sym-int1} one finds the following $\mathbb{Z}_2$-symmetry transformation 
    \begin{align}
     u\to 1/u \,,\qquad   \tilde{G}_{AB}\to u^{-2}\tilde G_{AB}
    \end{align}
    which is simply a consequence of the manifest $\Omega_1\leftrightarrow\Omega_2$ symmetry of the compactified Fefferman-Graham system.

Proposition~\bref{prop:diffeo} then implies that there exists a diffeomorphism realizing this $\mathbb{Z}_2$ symmetry. More precisely, using coordinates \eqref{twoscales-prop-u>0} the above tranformation of $u$ can be seen as a base space diffeomorphism. Note that the resulting symmetry transformation of the metric is a combination of the diffeomorphism and a Weyl transformation.
The $\mathbb{Z}_2$-symmetry is therefore realised as an involutive conformal isometry. In order to explicitly see how it works we introduce $v=1/u$ and observe the following identity:
    \begin{align}\label{Z_2-diffeo}
        \tilde G=\frac{du^2}{2u}+2u\gamma_{ab}(y)dy^ady^b= \frac{v}{2}\Bigg(d\Big(\frac{1}{v}\Big)\Bigg)^2+\frac{2}{v} \gamma_{ab}dy^ady^b=v^{-2}\Big(\frac{dv^2}{2v}+2v\gamma_{ab}dy^ady^b\Big)\,.
\end{align}

In Minkowski spacetime, this conformal isometry coincides with the standard inversion $X^A\to 2X^A/X^2$. More generally, it exchanges an interior null cone and null infinity, which is reminiscent of the Couch–Torrence inversion of extremal Reissner-Nordstr\"om black holes \cite{CouchTorrence1984}.

Getting back to generic solutions of the Fefferman-Graham system we have the following:
\begin{prop}
  An ambient metric $\tilde G$ satisfying the Fefferman-Graham system \eqref{gauge-fixed} for some function $u$  is regular as $u\to 0$ if and only if it is asymptotically flat as $u\to \infty$. 
\end{prop}

\begin{proof}
     By $\mathbb{Z}_2$-symmetry, regularity of $\tilde G$ as $u\to0$ implies regularity of $u^{-2}\tilde G$ as $u\to\infty$. But this is precisely the definition of asymptotic flatness of $\tilde G$ as $u\to\infty$.
\end{proof}

%%%%%%%%%%%

\subsection{Einstein from ambient Einstein}\label{sec: ads-ads}
%%%%%%%%%%%

An interesting feature of the two-scale system is that it can be used to obtain alternative, but equivalent, formulations of known systems. For instance, the Fefferman--Graham system \eqref{gauge-fixed} can be equivalently reformulated as a system describing an Einstein (rather than Ricci-flat) ambient space for an Einstein metric.

To see this, we follow Section~\bref{sec:equiv} and apply the following $GL(2,\mathbb R)$ transformation:
\begin{align}\label{transf-ads}
A=\begin{pmatrix}1&1/2\\1/2&1
\end{pmatrix}\notin\mathrm{Stab}(\lambda),\qquad \bar u=\frac{1+2u}{2+u}\,,\qquad  \bar G =\left(1+\frac{u}{2}\right)^{-2}\tilde G\,.
\end{align}
It follows from $0\leq u<\infty$ that $1/2\leq \bar u< 2$.
Equations \eqref{gauge-fixed} take the form
\begin{align}\label{ads-ambient}
    \begin{split}
        &\bar P_{AB}=\tfrac{1}{2}\bar G_{AB}\,,\\
        &\bar\nabla_A\bar\nabla_B\bar u=(\tfrac{5}{4}-\bar u)\bar G_{AB}\,,\\
        &\bar G^{AB}\partial_A\bar u\,\partial_B \bar u=(2-\bar u)(\bar u-1/2)\,.
    \end{split}
\end{align}
We refer to this system as the \textit{modified Fefferman-Graham system}. For a given solution $(\bar G,\bar u)$, the hypersurface $\bar u=1/2$ is null. It is the same curved null cone that is defined by $u=0$ in terms of the original fields $(\tilde G,u)$.

Given a solution $(\bar G,\bar u)$,
Proposition~\bref{prop-local-solution} implies that in the region where $1/2<\bar u<2$ there exists a coordinate system $(\bar u,y^a)$ such that $\bar G$ takes the form
\begin{align}\label{twoscales-ads-prop-u>0}
    \bar G=\frac{d\bar u^2}{(2-\bar u)(\bar u-1/2)}+(2-\bar u)(\bar u-1/2)\,\gamma\,,
\end{align}
where $\gamma=\gamma_{ab}(y)dy^a dy^b$ and $R_{ab}[\gamma]=\frac{9d}{16}\gamma_{ab}$, hence 
the hypersurfaces $\bar u=const$ are Einstein.

Although the modified Fefferman-Graham system is merely a field redefinition of the original ambient system \eqref{gauge-fixed}, it admits a different interpretation as a system describing the embedding of an Einstein space into an Einstein, rather than Ricci-flat, ambient space. In particular, it is worth spelling out explicitly the corresponding reformulation of Proposition~\bref{two-scales-prop-solution} adapted to the modified ambient system~\eqref{ads-ambient}. To this end, let $\gamma$
be an asymptotically AdS Einstein metric defined on a near-boundary coordinate patch. In a suitable coordinate system $(\sigma, x^m)$, it can be written as
\begin{align}
    \gamma=\frac{16}{9\sigma^2}\big(-d\sigma^2+\frac{1}{2}h_{mn}(\sigma,x)dx^mdx^n\big)\,.
\end{align}
Performing the coordinate change
\begin{align}
                t=\frac{\sqrt{2\bar u-1}}{\sigma\sqrt{2-\bar u}}\,,\qquad \rho=\sigma^2\,.
\end{align}
we can rewrite the metric $\bar G$ in \eqref{twoscales-ads-prop-u>0} as
\begin{align}\label{fef-gr-ads}
    \bar G=\left(1+\frac{\rho t^2}{2}\right)^{-2} \big(\,2\rho\, dt^2+2t\,dtd\rho+t^2h_{mn}(\sqrt{\rho},x)\,dx^mdx^n\big)\,.
\end{align}
This metric extends continuously to $\rho=0$, corresponding to $\bar u=1/2$. These formulas follow directly from the corresponding result for $\tilde G$ and the transformation \eqref{transf-ads}.

The modified Fefferman-Graham system defines a generalisation of the standard Fefferman--Graham construction that lifts a $d$-dimensional conformal metric $h$ to an Einstein, rather than Ricci-flat, $(d+2)$-dimensional ambient space. Namely, the metric \eqref{fef-gr-ads} is the corresponding analogue of the Fefferman--Graham ambient metric. Consequently, the standard formal existence and uniqueness results carry over to this setting via the transformation \eqref{transf-ads}.

\section{Compactified Spi/Ti system}\label{sec:ads-ambient}
%%%%%%%%%%%%%%%%%%%%%%%%%%%%%%%%%%%%
\begin{defi}\label{def-proj-ambient}
The two-scale system \eqref{two-scales} with 
\begin{align}\label{def-proj-ambient-lambda}
    \lambda=
    \begin{pmatrix}
        \epsilon & 0\\
        0 & 0
    \end{pmatrix},\quad\epsilon=\pm1\,,
\end{align}
and the Weyl densities subject to 
\begin{align}
    \Omega_i\geq0,\qquad i=1,2
\end{align}
is referred to as the compactified Spi system if $\epsilon=1$ and as the compactified Ti system if $\epsilon=-1$.
\end{defi}
Just like in the case of compactified Fefferman-Graham it is natural 
to consider Spi/Ti system on the manifold with corners such that there are two boundaries $\Omega_1=0,\Omega_2>0$ and $\Omega_2=0,\Omega_1>0$ and the corner $\Omega_1=\Omega_2=0$. Indeed, our regularity assumptions require $d\Omega_i$ to be non-vanishing near $\Omega_i=0$ and hence $\Omega_i$ can be considered boundary defining functions. From this perspective, the system with $\Omega_i>0$ is the one defined on the interior of this manifold while the one with  $\Omega_i \geq 0$ is the compactified one. As we are going to see, the corner $\mathcal C$ corresponds to asymptotic infinities of the metric $\gamma$ induced on the hypersurface ${\Omega_1}^{-1}\Omega_2=const \neq 0$.

Anticipating the discussion below, let us mention that for the Spi/Ti system the leaves $\Omega_2\propto\Omega_1\neq 0$ are isometric, up to a constant factor, to one and the same asymptotically flat hypersurface of signature $(p,q+1)$ if $\epsilon=+1$ and $(p+1,q)$ if $\epsilon=-1$. The corner  $\mathcal C$, which is the singular set of the canonical foliation, corresponds to the conformal null infinity of each leaf.
The second boundary $\Omega_1\neq0,\Omega_2=0$ corresponds to the extended spatial/temporal infinity of each leaf (aka Spi/Ti). Note that the latter (and related) terms will be understood for any signature with a slight abuse of terminology. The signatures in the general case follow from the analysis in Section~\bref{app: coord-corner}, while Table~\bref{tab:signatures} lists them for the cases relevant for applications in general relativity. 

\begin{table}[ht]
\centering

\renewcommand{\tabularxcolumn}[1]{m{#1}}
\setlength{\tabcolsep}{2.5pt}
\renewcommand{\arraystretch}{1.9}

\newcommand{\headercell}[1]{%
  \begin{minipage}[c][4.6em][c]{\linewidth}
    \centering
    #1
  \end{minipage}%
}

\newcommand{\tblmatrix}[4]{%
  \begingroup
  \renewcommand{\arraystretch}{1.1}%
  \begin{pmatrix}
    #1 & #2\\
    #3 & #4
  \end{pmatrix}%
  \endgroup
}

\begin{tabularx}{\linewidth}{
    >{\hsize=1.07\hsize
      \linewidth=\hsize
      \centering\arraybackslash}X
    |
    >{\centering\arraybackslash}m{2.05cm}
    |
    >{\hsize=0.93\hsize
      \linewidth=\hsize
      \centering\arraybackslash}X
    |
    >{\centering\arraybackslash}m{2.55cm}
    |
    >{\centering\arraybackslash}m{2.40cm}
}
\hline

\headercell{Matrix $\lambda$}
&
\headercell{Ambient space\\[2pt] \mbox{$\Omega_1\Omega_2\neq 0$}}
&
\headercell{Lorentzian hypersurface\\[2pt] \mbox{$\Omega_1=\Omega_2\neq0$}}
&
\headercell{Null hypersurface\\[2pt] \mbox{$\Omega_2=0,\;\Omega_1\neq0$}}
&
\headercell{Codim. two submanifold\\[2pt]
  \mbox{$\Omega_1=\Omega_2=0$}}
\\
\hline

$\tblmatrix{+1}{0}{0}{0}$
&
Einstein
$(2,d)$
&
Asymptotically flat $(1,d)$
&
Spi \qquad\qquad$(1,d-1,1)$
&
Scri\qquad\qquad
$(0,d-1,1)$
\\
\hline

$\tblmatrix{-1}{0}{0}{0}$
&
Einstein
$(1,d+1)$
&
Asymptotically flat
$(1,d)$
&
Ti \qquad\qquad$(0,d,1)$
&
Scri \qquad\qquad
$(0,d-1,1)$
\\
\hline

\end{tabularx}

\caption{\label{spi/ti signatures}Signature $(t,s)$ or inertial index $(t,s,n)$ of the (ambient or induced) metrics of various relevant loci in the two possible cases of compactified Spi/Ti ambient system which are relevant for physical applications to asymptotically flat spacetimes (i.e. $t$, $s$, $n$ are the respective number of timelike, spacelike, null directions).}

\label{tab:signatures}
\end{table}

%%%%%%%%%%%
\subsection{\texorpdfstring{Spi/Ti ambient system}{Spi/Ti ambient system}}
%%%%%%%%%%%
Just like in the case of compactified Fefferman-Graham system it is useful to begin the analysis with the simplified system for which $\Omega_1>0$. This can be seen as the system with boundary $\Omega_1=0,\Omega_2>0$ and  corner $\Omega_1=\Omega_2=0$ removed.
Substituting \eqref{def-proj-ambient-lambda} into \eqref{gauge-fixed1}, we obtain the following equations for $\tilde G_{AB} := \Omega_1^{-2} G_{AB}$ and $u:= {\Omega_1}^{-1}\Omega_2$:   
    \begin{align}\label{ads-ambient-fixed}
        \begin{split}
            &\tilde P_{AB}=-\tfrac12 \epsilon\,\tilde G_{AB},\\
            &\tilde\nabla_A\tilde\nabla_B u=\epsilon\,u\,\tilde G_{AB},\\
            &\tilde G^{AB}\partial_Au\,\partial_Bu=\epsilon\, u^2.
        \end{split}
    \end{align}
We call the above system a \textit{Spi/Ti ambient system} and assume it to be defined on the manifold with boundary at $u=0$. However, solutions defined on the interior $u>0$ only are also of interest.

Proposition~\bref{prop-local-solution} implies that the leaves of the canonical foliation $u=const\neq 0$ are Ricci-flat, that is to say  the pullback of $\tilde G$ to such a leaf has vanishing Ricci tensor. Drawing a parallel with the Fefferman-Graham system, the Spi/Ti ambient system can be regarded as an ambient space construction for a Ricci-flat space such that the ambient space is Einstein with the non-vanishing cosmological constant. Note that, just like for the ambient Fefferman-Graham system, the second equation implies that $V=\mathrm{grad}_{\tilde G}(u)$ is a conformal Killing vector field but now with a non-constant Weyl factor.

Given a solution to~\eqref{ads-ambient-fixed}, the corresponding canonical vector field $V=\mathrm{grad}_{\tilde G}(u)$ is tangent to the hypersurface $u=0$. Moreover, regularity conditions \eqref{regularity} ensure that $V$ is non-vanishing at $u=0$
while the second and third equations of \eqref{ads-ambient-fixed} say that $V$ restricted to this hypersurface is null and satisfies $ \mathcal{L}_V\tilde G=0$. In particular, $u=0$ is a conformal Killing horizon. Furthermore, the Levi-Civita connection of $\tilde G$ induces a connection on this horizon, as follows from the second equation in \eqref{ads-ambient-fixed}.

\begin{prop}\label{prop-strong-car}
    Let $(\tilde G,u)$ be a solution of \eqref{ads-ambient-fixed} and let $\mathcal N$ denote the hypersurface $u=0$.
Let also $h$ be the pullback of the ambient metric $\tilde G$ to $\mathcal N$, $\xi$ the restriction of the canonical vector field $V$ to $\mathcal N$, and $D$ the restriction of $\tilde \nabla$ to $\mathcal N$.  Then $(\mathcal N,h,\xi,D)$ is a strongly Carrollian geometry in the sense of \cite{Duval:2014uoa}, i.e. it satisfies
\begin{align}
    \ker h=\mathrm{span}\,\xi\,,\qquad D\xi=Dh=0\,,
\end{align}
and $D$ is torsion-free.
\end{prop}
\begin{proof}
Let us first show that $\ker h= \mathrm{span}\, \xi$. Indeed, any vector field $Y$ on $\mathcal{N}$ can be obtained as the restriction of a vector field $X$ defined on the ambient manifold and tangent to $\mathcal{N}$, that is, satisfying $(Xu)|_{u=0}=0$. This condition can equivalently be written as $\tilde{G}(V,X)|_{u=0}=0$, which immediately implies $h(\xi,Y)=0$. Notice that the kernel of $h$ cannot have dimension greater than one, since $\tilde{G}$ is non-degenerate and the codimension is one. 
Finally, the identities $Dh=0$ and $D\xi=0$ follow directly by restricting the relation $\tilde\nabla\tilde G=0$ to $\mathcal N$, together with the second equation in \eqref{ads-ambient-fixed}.
\end{proof}
The above proposition implies that the boundary $\cN$ is locally a fibre bundle over a $d$-dimensional (pseudo-)Riemannian manifold. Indeed, $\mathcal{L}_\xi h=0$ and $h(\xi,\cdot)=0$, so that $h$ is locally the pullback of a metric defined on the quotient of $\cN$ by the orbits of $\xi$. As we shall see, this quotient can be identified with the spatial/temporal infinity, in the sense of Ashtekar-Romano\footnote{We emphasise that in this article we do not consider spatial/timelike infinity in the framework of conformal compactification, where these infinities are represented by singular points.}~\cite{Ashtekar:1991vb},
 of the metric induced by $\tilde G$ on the hypersurface $u=1$. In this sense, $\cN$ is the direct analogue of the ``curved null cone'' in the ambient Fefferman-Graham system. 

By Proposition~\bref{prop-local-solution}, in the region $u>0$ one can introduce local coordinates $(u,y^a)$ in which the general solution of the system \eqref{ads-ambient-fixed} takes the form
\begin{align}\label{Spi-tilde-G}
    \tilde G=\epsilon\, \frac{du^2}{u^2}+u^2\gamma_{ab}(y)dy^ady^b
\end{align}
where $\gamma$ is an arbitrary Ricci-flat metric. Other way around, we may think that an arbitrary $(d+1)$-dimensional Ricci-flat metric $\gamma$ can be uplifted to the solution \eqref{Spi-tilde-G} of the system \eqref{ads-ambient-fixed} with $u>0$. 
In this sense, the metric \eqref{Spi-tilde-G} 
plays the role of an Einstein ambient metric $\tilde G$ for the Ricci-flat induced metric~$\gamma$.

\paragraph{Solutions smooth at $u=0$.}
%%%%%%%%%%%

We now discuss solutions of the Spi/Ti ambient system \eqref{ads-ambient-fixed} that are smooth at $u=0$.
According to Proposition~\bref{prop-local-solution0}, in a sufficiently small neighbourhood of any point on the hypersurface $u=0$, there exist local coordinates $(\rho,v,x^m)$ in which the general solution of the last two equations in \eqref{ads-ambient-fixed} takes the form
\begin{align}\label{smooth-sol-formula}
            \tilde G=\frac{1}{(1-\epsilon v\rho)^2}\Big(2dvd\rho+\epsilon\rho^2dv^2+h_{mn}(\rho,x)dx^mdx^n\Big)\,,\qquad
            u=\frac{\rho}{1-\epsilon v \rho}\,.
\end{align}
In the region $\rho>0$, the expression for $u$ can be regarded as a change of coordinates from $v$ to $u$. In the new coordinates $(\rho,u,x^m)$, the ambient metric takes the form
\begin{align}
    \tilde{G}=\epsilon\,\frac{du^2}{u^2}+u^2\left( -\,\epsilon\,\frac{d\rho^2}{\rho^4}+\frac{1}{\rho^2}\,h_{mn}(\rho,x)dx^mdx^n\right)\,.
\end{align}
Comparing this with \eqref{Spi-tilde-G}, we see that solutions of the system \eqref{ads-ambient-fixed} that are smooth at the boundary correspond to Ricci-flat metrics $\gamma$  which, in a neighbourhood of $\rho=0$, can be represented in the form
\begin{align}\label{gamma-smooth}
    \gamma=-\,\epsilon\,\frac{d\rho^2}{\rho^4}+\frac{1}{\rho^2}\,h_{mn}(\rho,x)dx^mdx^n\,.
\end{align} 
In the Riemannian setting, metrics of this type are known as scattering metrics in exact normal form \cite{melrose1994spectral}.

We are primarily interested in the case where the induced metric $\gamma$ has Lorentzian signature $(1,d)$. It is straightforward to check that such a metric is asymptotically flat at spatial infinity for $\epsilon=1$ and at temporal infinity for $\epsilon=-1$; see Definition~2.1 in \cite{Borthwick:2024skd} for a modern formulation of the original definition due to Ashtekar and Romano~\cite{Ashtekar:1991vb}. Moreover, the asymptotic boundary $\rho=0$, representing spatial/temporal infinity for $\gamma$, carries a naturally defined metric given by $h_{mn}(0,x)dx^mdx^n$. Note that, unlike the conformal metric induced at conformal infinity, this metric is uniquely defined up to an overall constant factor.

In the physics literature, asymptotic flatness at spatial or temporal infinity is usually formulated in terms of the existence of a Beig--Schmidt expansion near the corresponding infinity. In its simplest form, this expansion reads~\cite{Beig:1982ifu}
\begin{align}\label{Beig-Sch-exp}
\gamma'=-\,\epsilon\,\frac{\big(1+\sigma(x)\rho\big)^2}{\rho^4}\,d\rho^2+\frac{1}{\rho^2}\,h_{mn}(\rho,x)dx^mdx^n\,,
\end{align}
where $h_{mn}(\rho,x)$ is usually understood as a formal power series in the radial coordinate $\rho$, and $\sigma(x)$ is called the mass aspect. Comparing \eqref{gamma-smooth} with \eqref{Beig-Sch-exp}, we observe a key difference: starting from a solution $\tilde{G}$ that is smooth at the boundary, we obtain a vanishing mass aspect $\sigma(x)$. From a physical point of view, this is a very restrictive condition and, for $d=3$ and $\epsilon=1$, corresponding to spatial infinity, implies, in particular, that the ADM mass of the spacetime vanishes.
\begin{rem}
  The above argument for the vanishing of the mass aspect admits a natural geometric interpretation in terms of the so-called projective compactification. To see this, note that the metric $\gamma$ given by \eqref{gamma-smooth} is projectively compact of order $1$ in the sense of \cite{Cap:2013dva}. Moreover, any projectively compact Ricci-flat metric in a neighbourhood of the spatial/temporal projective boundary can be brought to this form.\footnote{The possibility of choosing coordinates and a defining function such that the coefficient of $d\rho^2$ has the form $\pm 1+\rho^2 f(\rho,x)$ was already established in \cite{Cap:2013dva}. Repeating the argument of \cite{Guillarmou:2019xray}, one can further achieve the coefficient $\pm1$.}
At the same time, projectively compact metrics are known~\cite{Borthwick:2024skd} to be asymptotically flat at spatial/temporal infinity, but the converse is not generally true: a non-trivial mass aspect $\sigma(x)$ provides an obstruction to the projective compactness of asymptotically flat metrics. See also the recent work \cite{Cap:2026projective}, which introduces the notion of projective pre-compactness, generalizing projective compactness to allow for a non-trivial mass aspect.
\end{rem}
Instead of studying the metrics induced on the hypersurface $u=1$ by the ambient metric $\tilde G$ solving the Spi/Ti ambient system, we now consider the converse question. Let $\gamma$ be a Ricci-flat metric of the form \eqref{gamma-smooth}. Then \eqref{smooth-sol-formula} gives, at least locally, a solution $(\tilde G,u)$ of the system \eqref{ads-ambient-fixed}. This solution extends smoothly to $u=0$ (or, equivalently, $\rho=0$). In the case of signature $(1,d)$, the induced metric defines a Carrollian extension of the spatial or temporal infinity of the metric $\gamma$. Recall that Proposition~\bref{prop-strong-car} states that the hypersurface $u=0$ carries the structure of a strongly Carrollian geometry.

Such Carrollian extensions of the corresponding infinities are well known in the literature and are referred to as Ti in the case of temporal infinity and Spi in the case of spatial infinity. This explains the name Spi/Ti ambient system for \eqref{ads-ambient-fixed}: it provides an ambient construction of Spi/Ti, directly analogous to the way the Fefferman-Graham ambient system describes the null cone. The notion of Spi/Ti has a long history and, in the case of spatial infinity, was first introduced in a different form by Ashtekar and Hansen~\cite{Ashtekar:1978zz}. The Carrollian interpretation of these structures was further developed in subsequent works. In particular, it was shown in \cite{Gibbons:2019zfs} that the structure introduced by Ashtekar and Hansen admits an interpretation as a weakly Carrollian manifold. Moreover, Ti and Spi for Minkowski spacetime were identified with homogeneous spaces of the Poincar\'e group in \cite{Figueroa-OFarrill:2021sxz}. More recently, a series of works \cite{Borthwick:2024skd,Borthwick:2023lye,Borthwick:2024wfn} investigated the Carrollian connection on extended infinities and the associated asymptotic symmetries using the Cartan geometry methods.

To summarise, given a smooth solution $(\tilde G,u)$ of the ambient system \eqref{ads-ambient-fixed} with signature according to Table \ref{spi/ti signatures}, we can identify
\begin{align}\label{N=Ti}
    \mathcal{N}= \begin{cases}
\mathrm{Ti}[\gamma]\,, & \epsilon=-1\,,\\
\mathrm{Spi}[\gamma]\,, & \epsilon=1\,,
\end{cases}
\end{align}
where $\gamma$ is the Ricci-flat metric \eqref{gamma-smooth} induced by $\tilde G$ on the hypersurface $u=1$ and $\mathcal{N}$ is the hypersurface $u=0$. It should be understood that, at least for the moment, $\mathrm{Spi}[\gamma]$, for instance, does not denote a rigorously defined Carrollian space associated with a manifold equipped with a Ricci-flat metric $\gamma$, but rather the outcome of one of the constructions available in the literature. An interesting possibility is to use the Spi/Ti ambient system to provide a constructive definition of Spi/Ti. However, for such a definition to be sufficiently general and well-defined, one needs to specify the admissible class of metrics $\gamma$ and the behaviour of $\tilde G$ as $u\to 0$. This requires further investigation and lies beyond the scope of the present work.

\begin{example}\label{ex-flat-ti-int}
Consider the $(d+1)$-dimensional Minkowski metric $\eta$ in coordinates adapted to the hyperbolic or de Sitter slicing:
\begin{align}\label{eta-slicing}
        \eta=-\epsilon\,\frac{d\rho^2}{\rho^4}+\frac{1}{\rho^2}\,h_\epsilon\,,
    \end{align}
where $h_{1}$ and $h_{-1}$ denote the metrics on de Sitter and hyperbolic spaces $dS_d$ and $H_d$, respectively, both of unit radius. The corresponding solution of the ambient system \eqref{ads-ambient-fixed} is
\begin{align}\label{flat-ti-spi}
                \tilde G=\frac{1}{(1-\epsilon v\rho)^2}(2dvd\rho+\epsilon\rho^2dv^2+h_\epsilon)\,,\qquad
            u=\frac{\rho}{1-\epsilon v \rho}\,.
\end{align}
It is straightforward to verify that the metric $\tilde G$ is conformally flat. Since its Schouten tensor is determined by \eqref{ads-ambient-fixed}, we conclude that $\tilde G$ is the metric of anti-de Sitter space $AdS_{d+2}$ for $\epsilon=-1$ and that of the $(d+2)$-dimensional pseudosphere with two timelike directions for $\epsilon=1$. Thus, in the flat case, $\mathrm{Ti}[\eta]$ and $\mathrm{Spi}[\eta]$ are realised as hypersurfaces in constant-curvature spaces of one higher dimension, equipped with the metric \eqref{flat-ti-spi}. This realisation is locally equivalent to the embeddings constructed in \cite{Figueroa-OFarrill:2021sxz}.
\end{example}

%%%%%%%%%%%
\paragraph{Solutions non-smooth at $u=0$.}
%%%%%%%%%%%
Earlier in this section, we saw that metrics that are asymptotically flat at spatial or temporal infinity and have vanishing mass aspect are locally in one-to-one correspondence with solutions of the system \eqref{ads-ambient-fixed} that are smooth at $u=0$. We now analyse solutions of \eqref{ads-ambient-fixed} arising from metrics $\gamma$ with non-vanishing mass aspect. It turns out that such solutions give rise to ambient metrics that develop singularities in their derivatives as $u$ approaches zero.
    \begin{prop}\label{propgamma-with-mass}
Let $\gamma$ be a $(d+1)$-dimensional Ricci-flat metric of the form 
     \begin{align}\label{gamma-with-mass}
        \gamma= \frac{(-\epsilon+ \nu(\rho,x)\rho)d\rho^2}{\rho^4}+\frac{1}{\rho^2}h_{mn}(\rho,x)dx^mdx^n\,.
    \end{align}
Then the corresponding solution $(\tilde G,u)$ given by \eqref{Spi-tilde-G} of the system \eqref{ads-ambient-fixed} in the region $u>0$ admits a continuous, but not necessarily smooth, extension to the boundary $u=0$. Moreover, the restriction of $(\tilde G,u)$ to $u=0$ is smooth.
    \end{prop}    
\begin{proof}The proof is somewhat technical and lengthy. It can be found in Appendix \bref{proofgamma-with-mass}.
\end{proof}
This gives another direct analogy between the Fefferman--Graham and the Spi/Ti ambient systems. Indeed, in both cases, smooth solutions of the corresponding ambient systems correspond to very special metrics $\gamma$. More general metrics, such as \eqref{gamma-with-mass}, lead to solutions that are continuous but not necessarily smooth at $u=0$. Note that Proposition~\bref{prop-strong-car}, which ensures that the hypersurface $u=0$ is equipped with a strongly Carrollian structure, assumes smoothness of the ambient solution $(\tilde G,u)$ and therefore does not directly apply in this more general setting. We leave this subtlety for future investigation.

%%%%%%%%%%%%%%%%%%%%%%%%%%%%%%%%%%%%%%%
\subsection{Ambient construction for Scri}
%%%%%%%%%%%%%%%%%%%%%%%%%%%%%%%%%%%%%%%

We now turn to the full Spi/Ti system, where both scales $\Omega_1$ and $\Omega_2$ are allowed to vanish. Recall that we assume this system to be defined on a manifold with corners and that $\Omega_1$ and $\Omega_2$ are the corresponding boundary defining functions. We have the following:
\begin{prop}\label{prop:corner=scri}
Let $(G,\Omega_1,\Omega_2)$ be a smooth solution of the compactified Spi/Ti system and let $\gamma$ be the metric induced by $\tilde G$ on the hypersurface $\Omega_1=\Omega_2 \neq 0$. This hypersurface is the interior of the submanifold $\Omega_1=\Omega_2$. Moreover, the boundary of the latter is the corner $\Omega_1=\Omega_2=0$, and the hypersurface $\Omega_1=\Omega_2$, equipped with the pullback of $G$, provides a conformal compactification of $\Omega_1=\Omega_2 \neq 0$. 
\end{prop}
\begin{proof}
Restricting $G$ to the hypersurface $\Omega_2=\Omega_1$, we recover $\Omega_1^2\gamma$ in the domain $\Omega_1>0$ by assumption. Furthermore, the pullback of $G$ to the hypersurface $\Omega_2=\Omega_1$ remains well-defined at the corner $\Omega_2=\Omega_1=0$.
\end{proof}

In other words, the Spi/Ti system provides not only an analogue of the Fefferman-Graham construction for Ricci-flat metrics and their spatial/temporal infinities, but also an ambient-like construction of Scri:
    \begin{align}
        \mathcal{C}=\mathrm{Scri}[\gamma]\,.
    \end{align}
Remarkably, the face $\Omega_2=0, \Omega_1>0$ is identified with Spi/Ti, so that Scri naturally arises as the asymptotic boundary of Spi/Ti. Let us recall, however, that all these statements strictly apply only to smooth solutions of the Spi/Ti system and hence to a rather restricted class of metrics. It is expected that this class can be extended to include physically relevant metrics at the price of allowing solutions that are not smooth at the boundaries. Although we leave a general analysis of this issue for future work, we consider a couple of particular cases below.

We begin with the flat case. 

\begin{example}\label{flat-spi}
Using the coordinate system introduced in Example~\bref{ex-flat-ti-int}, consider a solution of \eqref{ads-ambient-fixed} constructed from the flat metric $\gamma=\eta$, with $h_{\epsilon}$ written in the Poincar\'e patch:
\begin{align}
                        \tilde G=\frac{1}{(1-\epsilon v\rho)^2}\left(2dvd\rho+\epsilon\rho^2dv^2+z^{-2}(\epsilon dz^2-\delta_{\alpha\beta}d\sigma^\alpha d\sigma^\beta)\right)\,,\qquad
            u=\frac{\rho}{1-\epsilon v \rho}\,.
    \end{align}
In terms of a new coordinate system $(\Omega_1,\Omega_2,t,\sigma^\alpha)$, where
\begin{align}
    \Omega_1=z(1-\epsilon v\rho)\,,\qquad \Omega_2=z\rho\,,\qquad t=
    \epsilon z(v-\frac{\epsilon}{2}\rho v^2)\,.
\end{align}
one has
\begin{align}\label{ex:flat-comp-ti}
    \tilde G=\Omega_1^{-2}(\epsilon d\Omega_1^2+2\epsilon d\Omega_2dt-\delta_{\alpha\beta}d\sigma^\alpha d\sigma^\beta)\,.
\end{align}
It follows that the triple $(G=\Omega_1^2\tilde G,\Omega_1,\Omega_2)$ is a solution of the Spi/Ti system. Notice that the pullback of the metric $\Omega_2^{-2}G$ to $\Omega_1=0$ is simply the Minkowski metric in flat Bondi coordinates \cite{He:2019jjk,Donnay:2022wvx,Nguyen:2023vfz,Bekaert:2024tkv}, whereas the pullback of the metric $G$ to the corner corresponds to the metric on the associated Scri.  Figure~\bref{fig:minkowski} illustrates this example, with the compactified ambient space represented as a cylinder.
\begin{figure}[!htbp]
\centering
    \includegraphics[width=0.7\linewidth]{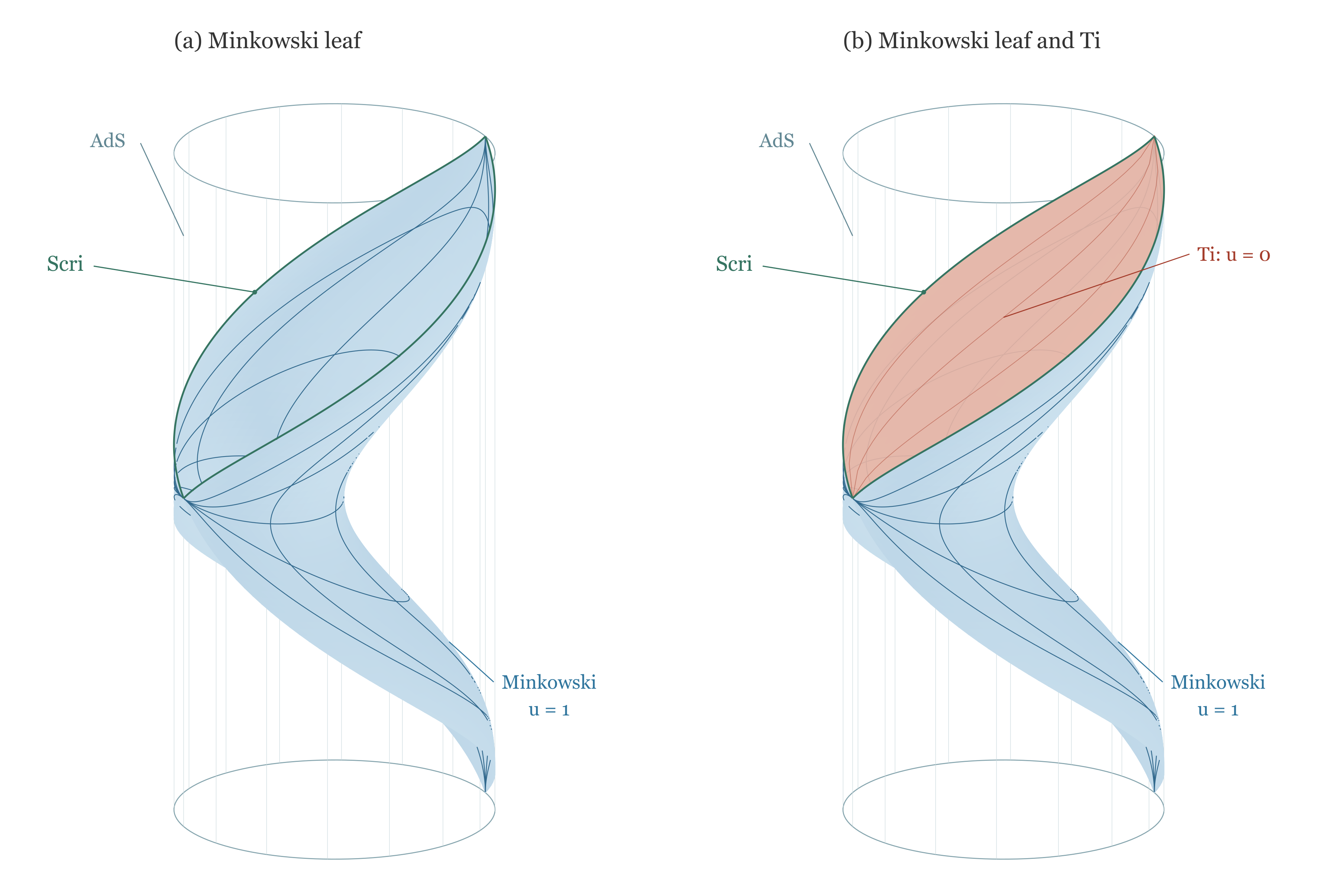}
    \caption{Ti ($\epsilon=-1)$ case for $\gamma=\eta$. The interior of the cylinder represents $\mathrm{AdS}_3$, while its boundary represents its conformal infinity. The interior of the blue two-dimensional surface represents Minkowski spacetime, and the intersection of its closure with the cylinder's boundary represents its conformal infinity, whose smooth part is Scri. The red surface represents Ti of the Minkowski leaf.}\
    \label{fig:minkowski}
\end{figure}

\end{example}

A slightly more general statement can be established in exactly the same way (compare with Proposition~\bref{prop:corner-coord}):
\begin{prop}
Suppose that a Ricci-flat metric $\gamma$ admits a coordinate system $(\rho,z,\sigma^\alpha)$, where $\alpha,\beta,\ldots$ take $d-1$ values,  in which
\begin{align}\label{gamma-com-corner}
    \gamma=-\epsilon \frac{d\rho^2}{\rho^4}+\frac{1}{z^2\rho^2}\left(\epsilon dz^2+k_{\alpha\beta}(\rho,z,\sigma)d\sigma^\alpha d\sigma^\beta\right).
\end{align}
Then the corresponding solution~\eqref{smooth-sol-formula},
defined for $\Omega_i>0$ and rewritten in terms of $(G,\Omega_1,\Omega_2)$, extends continuously to the manifold with corners, provided that
$k_{\alpha\beta}(\rho,z,\sigma)$ extends continuously to a neighbourhood
of $\rho=z=0$ and $\det k_{\alpha\beta}(0,0,\sigma)\neq0$. In the coordinate system $(\Omega_1,\Omega_2,t,\sigma^\alpha)$,  with $t>0,$ the corresponding solution takes the form
\begin{align}\label{G-com-corner}
        G=\epsilon\, d\Omega_1^2+2\epsilon\, d\Omega_2dt+k_{\alpha\beta}\Bigg(\frac{\Omega_2}{\sqrt{\Omega_1^2+2\Omega_2t}},\sqrt{\Omega_1^2+2\Omega_2t},\sigma\Bigg)d\sigma^\alpha d\sigma^\beta.
    \end{align}
In the interior region $\Omega_1>0$, $\Omega_2>0$, the triple $(G,\Omega_1,\Omega_2)$ is a smooth local solution of the compactified Spi/Ti system that extends continuously to the boundary faces and the corner.
\end{prop}
It follows from the above proposition that, for Ricci-flat metrics $\gamma$ of the form \eqref{gamma-com-corner}, the conformal infinity of Ti$[\gamma]$ or Spi$[\gamma]$ can be locally identified with the conformal infinity $\mathrm{Scri}[\gamma]$ itself.
Indeed, the pullback of \eqref{G-com-corner} to $\Omega_2=0$ can be interpreted as a conformal compactification of the boundary face $\mathrm{Ti}[\gamma]$ or $\mathrm{Spi}[\gamma]$, identified with the hypersurface $\mathcal{N}=\{\Omega_1>0$, $\Omega_2=0\}$ equipped with the metric $\Omega_1^{-2}G$. At the same time, Proposition~\bref{prop:corner=scri} implies that the corner $\Omega_1=\Omega_2=0$ can be identified with $\mathrm{Scri}[\gamma]$.

Note that, in view of \eqref{two-scales}, the restriction of the vector field $\mathrm{grad}_G\,\Omega_2$ to the hypersurface $\Omega_2=0$ is null and tangent to the boundary $\Omega_1=0$. Geometrically, this means that the compactification takes place in the ``horizontal'' direction, with Carrollian time regarded as vertical. For flat space, this identification can be verified directly using Example~\bref{flat-spi}. The idea that Scri may be interpreted as the asymptotic infinity of $\mathrm{Ti}$ or $\mathrm{Spi}$, albeit in a projective sense, was previously mentioned in \cite{Borthwick:2024wfn}. We emphasise again that the identification of Scri as an asymptotic boundary of Spi/Ti has been established only for the metrics of a very special form, in particular, those with vanishing mass aspect.

\subsection{Dual picture}\label{ti-dual}
%%%%%%%%%%%

According to Section~\bref{sec:equiv} a two-scale system corresponding to a given orbit of $GL(2,\fR)$ in the space of bilinear forms on $\fR^2$, has a family of equivalent formulations parametrised by particular $\lambda_{ij}$ belonging to the orbit. For instance, applying the $GL(2,\mathbb R)$ transformation to 
\begin{align}
A=\begin{pmatrix}0&1\\1&0
\end{pmatrix}\in GL(2,\mathbb R)
\end{align}
gives the standard $\lambda_{ij}$ \eqref{def-proj-ambient-lambda} of the Spi/Ti system, gives
\begin{equation}
\bar\lambda=\begin{pmatrix}
        0 & 0\\
        0 & \epsilon
    \end{pmatrix}
\end{equation}
which, in turn, determines an equivalent description of the 
same system. The corresponding 
ambient system reads as
\begin{align}\label{ads-ambient-omega2}
    \begin{split}
        &\bar P_{AB}=0\,,\\
        &\bar \nabla_{A}\bar \nabla_B\,\bar u=0\,,\\
        &\bar G^{AB}\partial_A\bar u\,\partial_B\,\bar u=\epsilon\,,
    \end{split}
\end{align}
where we changed the notations: $\tilde G \to \bar G$ and $u\to \bar u$. This system is of course equivalent to the standard Spi/Ti system~\eqref{ads-ambient-fixed} via field redefinition~\eqref{sym-int1} which in this case takes the following form:
\begin{equation}
\label{Gu-trans}
    \bar u=u^{-1}\,,\qquad  \bar G =u^{-2}\tilde G
\end{equation}
Let us now have a close look at system~\eqref{ads-ambient-omega2}. It describes a Ricci flat metric $\bar G$  together with its Killing vector $\bar V=\text{grad}_{\bar G} (\bar u)$ 
of squared norm $\epsilon$. In particular, in a suitable coordinate system $(\bar u,y^a)$, the general solution of \eqref{ads-ambient-omega2} takes the form
\begin{align}
    \bar G=\epsilon d\bar u^2+\gamma_{ab}(y)dy^ady^b\,,
\end{align}
where the induced metric $\gamma$ is Ricci-flat. In this dual picture, the canonical vector field $\bar V$ is parallel, the ambient space is Ricci-flat and foliated by spacelike or timelike hypersurfaces, $\bar u=\mathrm{const}$, which are themselves Ricci-flat. If the signature of the ambient metric is Lorentzian and the canonical vector field is timelike, then such spacetimes are sometimes called ultrastatic vacuum spacetimes in general relativity.

The limit $u\to 0$ considered in the previous subsections corresponds, in this case, to the limit $\bar u\to\infty$ for the conformally rescaled metric $\bar u^{-2}\bar G$. However, we showed in the preceding sections that this limit is well defined for metrics $\gamma$ belonging to a certain class. At the same time the dual ambient metric is Ricci-flat and hence it is natural to look at its null-infinity. We have the following:
\begin{prop}\label{spi=scri}
    Let $\gamma$ be a $(d+1)$-dimensional Ricci-flat metric of the form \eqref{gamma-smooth}. Then Spi or Ti associated with $\gamma$ via Proposition~\bref{propgamma-with-mass} can be identified with the conformal boundary of the dual ambient metric $\bar G=\epsilon\, d\bar u^2+\gamma$ with respect to the defining function $u=\bar u^{-1}$. 
\end{prop}
\begin{proof}
Passing from coordinate system $(\bar u,y^a)$ to $u,y^a$, with $u=\bar u^{-1}$ and considering the Weyl transformed  metric $\tilde G=u^2\bar G$ one finds that $u=0$ is precisely the null-infinity for $\bar G$ but, as we have seen,  it is precisely Spi/Ti for $\gamma$.
\end{proof}
In other words, somewhat informally we may write, say for $\epsilon=1$, that
\begin{align}\label{SpiScri}
    \mathrm{Spi}[\gamma]=\mathrm{Scri}_{\bar u^{-1}}[d\bar u^2+\gamma]\,,
\end{align}
where the subscript ${\bar u^{-1}}$ emphasises that the defining function of the asymptotic boundary of $\bar G$ is fixed to be $\bar u^{-1}$. Note that because the defining function is fixed, the degenerate metric induced on the asymptotic boundary is defined uniquely. This is in contrast to the possibly degenerate conformal structure induced by a bulk metric if a defining function is generic. 
Proposition~\bref{propgamma-with-mass} implies that  Proposition~\bref{spi=scri} admits an immediate generalisation to the metrics of the form \eqref{gamma-with-mass}, at the cost of the rescaled metric $\bar u^{-2}\bar G$ not being smooth at $\bar u^{-1}=0$.
\begin{example}
    We start with a flat solution of \eqref{ads-ambient-omega2},
    \begin{align}\label{bar-g-flat}
       \bar G= \epsilon\, d\bar u^2+\eta=\epsilon d\bar u^2-\epsilon\,\frac{d\rho^2}{\rho^4}+\frac{1}{\rho^2}\,h_\epsilon\,,
    \end{align}
    where the Minkowski metric $\eta$ is written in the de Sitter or hyperbolic slicing, depending on the sign of $\epsilon$; see \eqref{eta-slicing}.
The conformal rescaling $\bar u^{-2}\bar G$ takes the form
\begin{align}\label{bar-g-flat-2}
        \bar u^{-2}\bar G=\frac{1}{(1-\epsilon v\rho)^2}(2dvd\rho+\epsilon\rho^2dv^2+h_\epsilon)\,,
    \end{align}
where we use the new coordinate system where $\bar u$ is replaced by $v=\epsilon\left(\rho^{-1}-\bar u\right)$. This metric is smooth and nondegenerate at $\rho=0$. Furthermore, $\bar u^{-1}=\rho/(1-\epsilon v\rho)$ is a defining function for this hypersurface. Thus, $\rho=0$ is a conformal boundary of the metric $\bar G$ in \eqref{bar-g-flat}. The same metric \eqref{bar-g-flat-2} appeared in Example~\bref{ex-flat-ti-int}, where $\rho=0$ was identified with $\mathrm{Ti}$ or $\mathrm{Spi}$ of $\eta$, depending on the choice of $\epsilon$.
\end{example}

%%%%%%%%%%%%%%%%%%%%%%%%%%%%%%%%%%%%%
\section{Compactified Einstein degression system}\label{Einsteinsingularfoliation}
%%%%%%%%%%%%%%%%%%%%%%%%%%%%%%%%%%%%%

\begin{defi}
    We call the two-scale system \eqref{two-scales} with
    \begin{align}
  \lambda=      \epsilon\begin{pmatrix}
        1 & 0\\
        0 & 1
    \end{pmatrix},\quad\epsilon=\pm1\,,
    \end{align}
      the compactified Einstein degression system.
\end{defi}
By Corollary~\bref{cor:appA}, an arbitrary solution of the compactified Einstein degression system can be written, in a neighbourhood of the submanifold $\Omega_1=\Omega_2=0$, in the form
    \begin{align}\label{ambientmetsingEinst}
        G=\epsilon\delta^{ij}d\Omega_id\Omega_j+h_{mn}(\Omega_1^2+\Omega_2^2,x)dx^mdx^n\,.
    \end{align}

%%%%%%%%%%%

Consider the domain $\Omega_1\neq0.$ Substituting $\lambda_{ij}=\epsilon\, \delta_{ij}$ with $\epsilon=\pm 1$ into \eqref{gauge-fixed1}, we obtain the following equations for the ambient metric $\tilde G$ and the level function $u$:
\begin{align}\label{singularfoliation}
\begin{split}
        &\tilde P_{AB}=-\tfrac12\epsilon\,\tilde G_{AB},\\
        &\tilde \nabla_A\tilde \nabla_Bu=\epsilon\, u\,\tilde G_{AB},\\
        &\tilde G^{AB} \partial_Au\,\partial_Bu=\epsilon(u^2+1)\,,
\end{split}
\end{align}
One can choose local coordinates $(u,y^a)$ in which the corresponding ambient metric takes the form
\begin{align}\label{generic-tilde-last}
    \tilde G=\epsilon\,\frac{du^2}{u^2+1}
+(u^2+1)\,\gamma_{ab}(y)\,dy^a\,dy^b\,,
\end{align}
where $\gamma=\gamma_{ab}(y)\,dy^a\,dy^b$ is a $(d+1)$-dimensional Einstein metric
whose Ricci tensor satisfies
\begin{align}
        R_{ab}[\gamma]\,=\,-\,\epsilon\,d\,\gamma_{ab}\,.
    \end{align}
Note that thanks to the last equation of \eqref{singularfoliation} the canonical vector field is everywhere timelike (respectively, spacelike) for $\epsilon=1$ (respectively, for $\epsilon=-1$) and that the ambient metric \eqref{generic-tilde-last} is regular for all real values of the coordinate $u$.

Thus, locally, this system has the structure of an Einstein warped product, foliated by totally umbilic Einstein hypersurfaces. We can bring the metric \eqref{generic-tilde-last} to a more standard form by performing the coordinate transformation $u=\sinh z$. Then
\begin{align}
\tilde G=\epsilon dz^2+\cosh ^2(z)\,\gamma_{ab}(y)\,dy^a\,dy^b\,.
\end{align}
When $\gamma$ has constant curvature, this metric is known in the context of AdS degression \cite{Metsaev:2000qb,Artsukevich:2008vy}, where fields on AdS${}_d$ are described in terms of fields defined on the ambient AdS${}_{d+1}$ space. The term Einstein degression reflects the fact that this geometry generalises the AdS slicing underlying dimensional degression to Einstein metrics of nonconstant curvature.

Let us return to the full two-scale system. In complete analogy with Proposition~\bref{prop:corner=scri}, one can show that the singular set $\Omega_1=\Omega_2=0$ of the canonical singular foliation corresponds to the asymptotic infinity of the Einstein metric $\gamma$ in \eqref{generic-tilde-last}. The submanifold $\Omega_1=\Omega_2=0$ carries a conformal structure with representative metric $h_{mn}(0,x)dx^mdx^n$ of signature $(p-1,q+1)$ for $\epsilon=1$ and signature $(p+1,q-1)$ for $\epsilon=-1$, obtained as the pullback of the ambient metric \eqref{ambientmetsingEinst}.

%%%%%%%%%%%%%%%%%%%%%%%%%%%%%%%%%%%%%%
\section{Compactified Brinkmann system}\label{CompactifiedBrinkmannspace}
%%%%%%%%%%%%%%%%%%%%%%%%%%%%%%%%%%%%%%

\begin{defi}\label{CompactifiedBrinkmannsp}
    We call the two-scale system \eqref{two-scales} with a vanishing matrix $\lambda=0$
    the compactified Brinkmann system.
\end{defi}
For this choice of the matrix $\lambda$, we obviously have $
Stab(\lambda)=Stab(0)
=GL(2,\mathbb{R}).$

\begin{rem}\label{remBargmann}
    For the two-scale system \eqref{two-scales} to admit solutions satisfying the regularity conditions \eqref{regularity} at $\Omega_1=\Omega_2=0$, the signature of $G_{AB}$ must be $(p+1,q+1)$ with $p\geq1$ and $q\geq1$. In particular, the ambient metric cannot be Lorentzian if the solution is to be well-defined at the submanifold $\Omega_1=\Omega_2=0$.  Indeed,  $\lambda=0$ and the second line of \eqref{two-scales} imply that the gradient vector fields $\mathrm{grad_G}(\Omega_i)$ are null and orthogonal to each other at $\Omega_1=\Omega_2=0$. Lemma \bref{lemma-rho} and the second line of \eqref{two-scales-without-rho} then imply that the same property holds in a neighbourhood. On the other hand, it is well known that the existence of two linearly independent, mutually orthogonal null vectors requires at least two timelike and two spacelike directions.
\end{rem}

%%%%%%%%%%%
\subsection{\texorpdfstring{$\Omega_1=\Omega_2=0$ submanifold}{Omega1=Omega2=0 submanifold}}
%%%%%%%%%%%

We begin our analysis of the compactified Brinkmann system by examining the structures induced on the singular set $\mathcal C$. The following proposition applies to metrics $G$ of signature $(p+1,q+1)$ with $p\geq1$ and $q\geq1$:
\begin{prop}\label{prop:2-carr}
    Let $(G,\Omega_1,\Omega_2)$ be a solution of the compactified Brinkmann system. Then the codimension-two submanifold $\mathcal{C}=\{\Omega_1=\Omega_2=0\}$ is endowed with:
    \begin{itemize}
        \item a degenerate conformal metric of signature $(p-1,q-1)$, and
        \item two vector fields of Weyl weight minus one that span its kernel. Moreover, these vector fields generate an integrable two-dimensional distribution.
    \end{itemize}
\end{prop}
\begin{proof}
With $\lambda_{ij}=0$, the second line of \eqref{two-scales} implies that ${\mathrm{grad}_G\,\Omega_1}$ and ${\mathrm{grad}_G\,\Omega_2}$ are null and orthogonal to each other on the submanifold $\mathcal{C}$. 
The nondegeneracy condition $d\Omega_1\wedge d\Omega_2|_{\mathcal{C}}\neq0$ guarantees that the distribution they span is two-dimensional.
The fact that both vector fields belong to the kernel of the pullback of $G$ to $\mathcal{C}$ can be established by an argument analogous to the proof of Proposition~\bref{prop-strong-car}. Finally,
the integrability of the corresponding distribution on $\mathcal C$ follows from \eqref{U-comm}.
\end{proof}

Notice that the structures on $\mathcal C$ described in the above proposition naturally motivate a generalisation of conformal Carrollian geometry in which the degenerate conformal metric has a two-dimensional integrable kernel distribution. We also draw attention to \cite{Figueroa-OFarrill:2021sxz}, where the Poincar\'e homogeneous space Ni, likewise equipped with a degenerate metric with a two-dimensional kernel, was introduced.

%%%%%%%%%%%
\subsection{\texorpdfstring{$\Omega_1\neq 0$}{Omega non 0} domain}
%%%%%%%%%%%

Consider the domain $\Omega_1\neq0.$ Substituting $\lambda_{ij}=0$ into \eqref{gauge-fixed1}, we obtain the following equations for the ambient metric $\tilde G$ and the level function $u$:   
\begin{align}\label{Bargmann}
        \begin{split}
            &\tilde{P}_{AB}=0,\\
            &\tilde \nabla_A\tilde \nabla_B u=0,\\
            &\tilde G^{AB}\partial_Au\,\partial_Bu=0.\\
        \end{split}
\end{align}
Applying the general formula \eqref{sym-int1} to the present case, we find that the system \eqref{Bargmann}  possesses the following symmetries
\begin{align}
    u\to u'=\frac{A_{2}{}^{1}+A_{2}{}^{2}u}{A_{1}{}^{1}+A_{1}{}^{2}u},\qquad \tilde{G}\to\tilde G'=(A_{1}{}^{1}+A_{1}{}^{2}u)^{-2}\tilde G\,,
\end{align}
for any $A\in GL(2,\mathbb R)$ provided that $A_{1}{}^{1}+A_{1}{}^{2}u\neq0$.

A pseudo-Riemannian manifold endowed with a null parallel vector field $V$ is often called a Brinkmann space, thereby the choice of terminology in Definition \bref{CompactifiedBrinkmannsp}. Here, the  canonical vector field $V$ is the gradient of $u$, so the second and third line in \eqref{Bargmann} are precisely the local defining conditions for a Brinkmann space. 
A Bargmann manifold is a Lorentzian manifold endowed with a nowhere-vanishing null parallel vector field \cite{Duval:1984cj}. In this sense, a Bargmann manifold is a particular type of Brinkmann space of Lorentzian signature, i.e. $(p+1,q+1)=(1,d+1)$. The assumption that the canonical vector field $V$ is nowhere vanishing outside $\mathcal{C}$ already follows from the regularity conditions \eqref{regularity}.
It is well-known that Bargmann manifolds play, for Newton--Cartan and Carrollian geometries, a role analogous to that played by the Fefferman--Graham construction in conformal geometry in the following sense. First, a Newton--Cartan geometry is obtained on the quotient of $M|_{\Omega_1\neq 0}$ by the null congruence generated by the gradient of $u$ \cite{Duval:1984cj}. Second, a Carrollian geometry is induced on each null hypersurface $u=\mathrm{const}$, see \cite{Duval:2014uoa} for details. It should be emphasised, however, that Ricci-flatness is not part of the standard definition of a Bargmann manifold and, in general, imposes non-trivial restrictions on the induced Newton--Cartan and Carrollian geometries. More precisely, the induced Newton–Cartan connection on the former satisfies precisely the vacuum Newtonian field equations (cf. \cite{Duval:1984cj}) while the latter obey the null characteristic constraint equations on the Carrollian leaves (see, e.g., \cite{Mars:1993mj}). Note that, in this case, the coordinate system introduced in Proposition~\bref{prop-local-solution0} coincides with Fermi-Brinkman coordinates \cite{Brinkmann:1925fr}
\begin{align}
            \tilde G=2dudv+h_{mn}(u,x)dx^mdx^n\,,
\end{align}
where $h=h_{mn}dx^mdx^n$ are 
metrics on the leaves $u=\mathrm{const}$. 

Given a solution $(G,\Omega_1,\Omega_2)$ of the compactified Brinkmann system, the corner $\mathcal{C}$ arising in Proposition~\bref{prop:2-carr} can be interpreted as the conformal infinity of any null hypersurface $u=\mathrm{const}$ or, equivalently, in the region $\Omega_1>0$, on the null hypersurface $\Omega_2=\mathrm{const}\cdot\Omega_1$. Indeed, if $\gamma$ denotes the pullback of $\tilde G$ to $u=1$, then the conformal metric induced by $G$ on $u=1$ is given by $\Omega_1^2\gamma$ and the limit $\Omega_1\to0$ corresponds precisely to the corner $\mathcal{C}$.

%%%%%%%%%%%%%%%%%%%%%%%%%%%%%%%%
\section{Fields on a background}\label{sec: fields}
%%%%%%%%%%%%%%%%%%%%%%%%%%%%%%%%

We may consider additional fields propagating on the background of the two-scale system. A relatively simple example of such fields is provided by conformal fields, in the sense of Weyl-covariant fields propagating on a background which is a conformal manifold.

\subsection{General remarks}

Let $\Phi$ be a field on $M$ (possibly gauge) of Weyl weight $w$  such that its equations of motion 
$E[\Phi;G]=0$
are Weyl-invariant, where $G$ denotes the dependence on the background conformal metric. 
Such fields, since their equations of motion do not depend on $\Omega_i$ at all, inherit the $Stab(\lambda)$-symmetries (see Proposition~\bref{prop:stabilizer}) of the corresponding two-scale system. In particular,
   in the compactified Fefferman--Graham setup, they are invariant under the $\mathbb{Z}_2$ symmetry. Repeating the arguments from Section~\bref{sec:Z2}, it is easy to see that the asymptotic expansions of these fields near $\Omega_2=0$ and $\Omega_1=0$ (interpreted in the flat case as the null cone and null infinity, respectively) are related. This leads to a curved (for $d>3)$ and coordinate-free analogue of the Cone $\leftrightarrow$ Scri duality discussed in \cite{Bekaert:2024tkv}.

We note that this fact depends crucially on starting from equations $E[\Phi;G]$ whose dependence on the background sector is only through the conformal metric. Of course, one may consider a much broader class of Weyl-invariant equations depending also on $\Omega_i$ In particular, any diffeomorphism-invariant equation on the interior of $M$ can be lifted to a Weyl-invariant equation on the whole $M$ via the St\"uckelberg trick, see e.g.~\cite{Deser:1970hs, Gover:2008pt,Grigoriev:2025gvk}.

%%%%%%%%%%%
\subsection{Ambient description of massive scalar fields}
%%%%%%%%%%%
We start with a two-scale system \eqref{two-scales},
together with a conformal scalar field $\Phi$ of Weyl weight $-\frac{d}{2}$, subject to the following equations:
\begin{align}\label{conf-scalar}
\begin{split}
    &\big(\nabla^2-\tfrac{d}{2}\,P\big)\Phi=0,\\
    &\big(V^A\partial_A+\tfrac{d}{2}\,(\Omega_2\rho_1-\Omega_1\rho_2)-\kappa\big)\Phi=0\,,
\end{split}
\end{align}
where $V=V^A\partial_A$ is the canonical vector field and $\kappa$ is a constant. 
The first equation in \eqref{conf-scalar} is the Yamabe (or conformally coupled scalar) equation. The second equation in \eqref{conf-scalar} can be written as $\mathcal{D}_V\Phi=\kappa \Phi$, where $\mathcal{D}_V$ is the first-order differential operator introduced in \cite{eastwood2007symmetries}. There it was also shown  that, for any conformal Killing vector $V$, the operator $\mathcal{D}_V$ is a symmetry of the Yamabe operator, so the system \eqref{conf-scalar} is compatible.
In fact, $\mathcal{D}_V$ is nothing but the infinitesimal transformation of the density $\Phi$ under the conformal Killing vector field $V$ satisfying \eqref{prop-conf-killing-eq}. The second equation in \eqref{conf-scalar} can be interpreted as a homogeneity condition with respect to this conformal isometry. 
\begin{rem}
    Note that the symmetries of the two-scale system given in Proposition~\bref{prop:stabilizer} are not, in general, symmetries of the second equation in \eqref{conf-scalar}. Nevertheless, this equation transforms covariantly, yielding an equation of the same form with
\begin{align}
    \kappa'=\det(A)\,\kappa\,.
\end{align}
\end{rem}

In the domain $\Omega_1>0$, we introduce
the Weyl-invariant scalar field $\tilde\Phi=\Omega_1^{d/2}\Phi$ and the equations \eqref{conf-scalar} take the form
\begin{align}\label{systemphitilde}
    \begin{split}
        &\Big(\tilde \nabla^2+\lambda_{11}\frac{d(d+2)}{4}\Big)\tilde\Phi=0\,,\\
&\Big(\tilde G^{AB}\partial_Au\,\partial_B
+\frac d2(u\lambda_{11}-\lambda_{12})-\kappa\Big)\tilde\Phi=0.    \end{split}
\end{align}
For Bargmann spaces ($\lambda=0$), the quotient by $V$ carries a Newton-Cartan geometry. For $\kappa=im$, these equations reduce to the Schrödinger equation on this quotient~\cite{Duval:1984cj}.

In the coordinate system \eqref{generic-tilde-g} of Proposition~\bref{prop-local-solution}, the second equation in \eqref{systemphitilde} reduces to an ordinary differential equation, allowing the system to be solved in full generality.
\begin{prop}
In the region where $\Omega_1\neq0$ and $P_2(u)\neq0$, the general solution of \eqref{systemphitilde}, expressed in the coordinate system \eqref{generic-tilde-g}, takes the form 
\begin{align}
    \tilde\Phi(u,y)=|P_2(u)|^{-d/4}e^{\kappa I(u)}\phi(y)\,,
    \end{align}
    where  $I(u)=\int \frac{du}{P_2(u)}$ is any primitive of the inverse of the polynomial $P_2$ while $\phi(y)$ is a function on a leaf $u=\mathrm{const}$ which is solution of the following Helmholtz (or Klein-Gordon) equation  
    \begin{align}
        \left(
\nabla^2_\gamma+\sgn(P_2)\Big(\kappa^2
+\det\lambda\,\frac{d^2}{4}\Big)
\right)\phi=0\,. 
    \end{align}
\end{prop}
As for the two-scale system itself, such solutions are generally singular at $P_2(u)=0$. The subclass of solutions that remain regular there singles out fields with a particular asymptotic behaviour.
\begin{example}
    If we take as a background a fixed solution to Fefferman-Graham system~\eqref{gauge-fixed} then in the region $u>0$ equations \eqref{systemphitilde} take the following form:
    \begin{align}\label{FGscalar}
        \tilde\nabla^2\tilde\Phi=0\,,\qquad (\tilde G^{AB}\partial_Au\,\partial_B+\Delta)\tilde\Phi=0 
    \end{align}
    with $\Delta:= \frac{d}{2}-\kappa$.
These equations are the standard ambient scalar equations (see e.g. \cite{Biswas:2002nk} for the flat case and \cite{Bekaert:2017bpy} for its curved analogue) corresponding to the unique lift of a massive scalar field defined on the hypersurface $u=\mathrm{const}$ to the ambient space. In fact, using the coordinate system \eqref{twoscales-prop-u>0}, the general solution in the $u>0$ region is
    \begin{align}\label{phi-fg-gen}
        \tilde\Phi(u,y)=u^{-\Delta/2}\phi(y),\qquad \big(\nabla^2_\gamma+\Delta(\Delta-d)\big)\phi(y)=0\,.
    \end{align}
    The background $\mathbb{Z}_2$ symmetry $u\to u^{-1}$ discussed in Section~\bref{sec:Z2} extends  naturally to the scalar field $\tilde \Phi$ as $\tilde\Phi\to u^{d/2}\tilde\Phi$. As follows from \eqref{phi-fg-gen}, this transformation is not, strictly speaking, a symmetry, but instead it gives us a solution of \eqref{FGscalar} with $\Delta\to d-\Delta$. 
    
Under the assumptions of Proposition~\bref{two-scales-prop-solution} we can employ Fefferman--Graham coordinate system $(\rho,t,x^m)$ related via \eqref{coord-t-rho}, which is regular at $u=0$ or, equivalently, $\rho=0$ where  \eqref{phi-fg-gen} takes the following form:
\begin{align}
        \tilde\Phi(\rho,t,x)=t^{-\Delta}\rho^{-\Delta/2}\phi(\rho,x)\,.
    \end{align}
It follows, the smoothness of $\tilde \Phi$ at $\rho=0$ requires $\phi(\rho,x)=\rho^{\Delta/2}\phi_{reg}(\rho,x)$ for some smooth $\phi_{reg}(\rho,x)$. The boundary value $\varphi$ of the ambient scalar field $\tilde\Phi$ at $\rho=0$ then takes the form 
    \begin{align}
        \varphi(t,x):=\tilde\Phi(0,t,x)=t^{-\Delta}\phi_{reg}(0,x)   \,,
    \end{align}
 i.e., $\varphi$ is a homogeneous function  of degree $-\Delta$.
    
\end{example}
\begin{example}
Take as a background a fixed solution of the Spi/Ti ambient system defined in~\bref{def-proj-ambient}. In the region $u>0$ equations \eqref{systemphitilde} take the following form:
 \begin{align}
            \Big(\tilde\nabla^2+\epsilon\frac{d(d+2)}{4}\Big)\tilde\Phi=0\,,\qquad (\tilde G^{AB}\partial_Au\,\partial_B+\epsilon\,\tfrac{d}{2}\,u-\kappa)\tilde\Phi=0.
        \end{align}
        The general solution in the $u>0$ region reads as
        \begin{align}
\tilde\Phi(u,y)=
u^{-d/2}e^{-\epsilon\kappa/u}\phi(y),
\qquad
\left(
\nabla^2_\gamma+\epsilon\kappa^2
\right)\phi(y)=0.
    \end{align}
From the second equation, the physical mass is given by $m^2=\epsilon\kappa^2$. Restricting $\gamma$ to be of the form \eqref{gamma-smooth}, we pass to the coordinates $(\rho,v,x)$ that are regular at $u=0$, related to the original coordinates by $u=\frac{\rho}{1-\epsilon \rho v}$. The general solution takes the form
\begin{align}
    \tilde\Phi(\rho,v,x)=\rho^{-d/2}(1-\epsilon v\rho)^{d/2}e^{\kappa v-\epsilon\kappa/\rho}\phi(\rho,x)\,.
\end{align}
Consequently, smoothness at $\rho=0$ $(u=0)$ requires 
\begin{align}
    \phi(\rho,x)=\rho^{d/2}e^{\epsilon\kappa/\rho}\phi_{reg}(\rho,x)\,,
\end{align}
where $\phi_{reg}(\rho,x)$ is smooth at $\rho=0$. Thus, in the Ti-ambient case, $\kappa=\pm im$ corresponds to the oscillatory behaviour at temporal infinity, while in the Spi-ambient case, $\kappa=-m$  corresponds to a decaying behaviour at spatial infinity. The boundary value of the ambient scalar field $\tilde\Phi$ then takes the form
\begin{align}
  \varphi(v,x):=  \tilde\Phi(0,v,x)=e^{\kappa v}\phi_{reg}(0,x)\,.
    \end{align}
For $\epsilon=-1$ and $\kappa=\pm im$, these are precisely the scattering data induced by a massive field on Ti, as studied in detail in \cite{Borthwick:2024skd}.

For Ricci-flat metrics $\gamma$ with a non-vanishing mass aspect \eqref{gamma-with-mass}, it is natural to pass to coordinates \eqref{change-mass} in which the metric extends continuously to $\rho=0$. Then, for the ambient scalar field to admit a continuous limit at $\rho=0$, the asymptotic behaviour of $\phi(\rho,x)$ must receive an additional correction corresponding to the mass aspect of $\gamma$:
\begin{align}
    \phi(\rho,x)=\rho^{d/2+\nu(0,x)\kappa/2}e^{\epsilon\kappa/\rho}\phi_{reg}(\rho,x)\,.
\end{align}
\end{example}

\section*{Acknowledgments}

We thank Y. Herfray for discussions.  M.M. also acknowledges fruitful discussions
with Jack Borthwick, Ivan Dneprov and Viacheslav Krivorol. The work of X.B. was supported in part by the project ScattHoloGR, ANR-25-CE40-4883, France. The work of M.G. and M.M. was supported by the ULYSSE Incentive
Grant for Mobility in Scientific Research [MISU] F.6003.24, F.R.S.-FNRS, Belgium.

\appendix

%%%%%%%%%%%%%%%%%%%%%%%%%%%%%%%
\section{Some technical proofs}
%%%%%%%%%%%%%%%%%%%%%%%%%%%%%%%

This appendix contains various lengthy technical proofs which have been moved here in order to improve the flow of the discussion in the body of the paper.

\subsection{Proof of Proposition \bref{prop-local-solution0}}\label{proofPropproplocal0}
\begin{proof}
It was shown in Theorem 3.1 of \cite{catalano2006closed} that any pseudo-Riemannian manifold equipped with a closed conformal Killing vector $V$ admits, in a neighbourhood of any point $p$ such that $V|_p\neq0$, coordinates $(y,v,x^m)$ in which the metric takes the form
\begin{align}
    \tilde G=-a(y)dv^2+2dy d v+g_{mn}(y,v,x)dx^mdx^n\,,
    \end{align}
where the metric $g=g_{mn}dx^mdx^n$ satisfies
    \begin{align}\label{catalano-proof1}
    \big(\partial_v +a(y)\partial_y-a'(y)\big) g_{mn}=0\,.
\end{align}
The canonical vector field $V$ in these coordinates takes the form
\begin{align}
    V=\frac{\partial}{\partial v}+a(y)\frac{\partial}{\partial y}\,.
\end{align}
We take one further step and straighten the vector field $V$ by choosing a function $\rho(y,v)$ satisfying
\begin{align}
\mathcal{L}_V\rho=0\,,\qquad \rho(y,0)=y+c
\end{align}
and using it as one of the coordinates in place of $y$. Here, $c$ is a constant that will be convenient to fix later. 
The inverse transformation  $y=Y(\rho,v)$ is locally well defined and is determined by the flow equations
\begin{align}
    \partial_vY=a(Y)\,,\qquad Y(\rho,0)=\rho-c\,.
\end{align}
In terms of the new coordinates, equation \eqref{catalano-proof1} is solved by
\begin{align}
    g_{mn}\big(Y(\rho,v),v,x\big)=\partial_\rho Y(\rho,v)h_{mn}(\rho,x)\,.
\end{align}
Taking $V=\tilde\nabla u$ and using \eqref{gauge-fixed1}, we immediately obtain
\begin{align}
\begin{split}
&a(Y)=P_2(u)\,,\\
&    du=dY\quad \Rightarrow u=Y+\mathrm{const}.
    \end{split}
\end{align}
Using the constant $c$ introduced above to eliminate this freedom, we obtain the claimed result.
\end{proof}

\subsection{Proof of Proposition \bref{prop-local-solution}}\label{proofPropproplocal}

By the regularity assumptions \eqref{regularity}, $du\neq0$, and hence $u$ can be chosen as one of the local coordinates. Moreover, by the third equation in \eqref{gauge-fixed1} we have
\begin{align}
    V(u)=\tilde G^{AB}\partial_Au\partial_Bu=\tilde G^{uu}=P_2(u)\,.
\end{align}
 Since $P_2(u)\neq0$ in the region under consideration, $V$ is transverse to the level hypersurfaces of $u$. Therefore, the remaining coordinates $y^a$ can be chosen locally to be constant along the integral curves of $V$. Hence,
\begin{align}
    0=V(y^a)=\tilde G^{BC}\partial _Bu\partial_Cy^a=\tilde G^{ua}
\end{align}
and $V=\tilde G^{uA}\partial_A=P_2(u)\partial_u$. It follows that $\tilde G_{ua}=0$ and $\tilde G_{uu}=1/{P_2(u)}$. Considering the $ab$ components of the second equation in \eqref{gauge-fixed1}, we obtain 
\begin{align}
    \frac{P{_2}(u)}{2}\partial_u\tilde G_{ab}=\frac{P'_2(u)}{2}\tilde G_{ab} \quad \Longrightarrow\quad \tilde G_{ab}=|P_2(u)|\gamma_{ab}(y)\,,
\end{align}
where the sign of $P_2$ has been absorbed into the definition of $\gamma$.
The result can then be substituted into the standard curvature formulas for warped-product metrics to express the Ricci tensor of $\tilde G$ in terms of the Ricci tensor of $\gamma$:
    \begin{align}
\tilde R_{ab}=R_{ab}[\gamma]
-\sgn(P_2)
\left(
\frac12P_2P_2''
+\frac d4(P_2')^2
\right)\gamma_{ab}\,.
    \end{align}
Using the first equation in \eqref{gauge-fixed1} and the explicit form of $P_2$, we obtain
$R_{ab}[\gamma]=-d\,\sgn(P_2)\det\lambda\,\gamma_{ab}$. 
%%%%%%%%%%%
\subsection{Proof of Lemma \bref{lemma-rho}}\label{app: rho-gauge}
Under Weyl transformations $G\to e^{2\alpha(x)}G$, the functions $\rho_i$ transform as \cite{BEG}
\begin{align}
    \rho_i\to e^{-\alpha}(\rho_i-G^{AB}\partial_A\Omega_i\,\partial_B\alpha-\frac{\Omega_i}{2}G^{AB}\partial_A\alpha\,\partial_B\alpha).
\end{align}
Thus, the problem of gauge fixing reduces to finding a function $\alpha$ satisfying the two equations
\begin{align}\label{rho-fix} G^{AB}\partial_A\Omega_i\,\partial_B\alpha+\frac{\Omega_i}{2}G^{AB}\partial_A\alpha\,\partial_B\alpha=\rho_i\,,\qquad i=1,2\,.
\end{align}
Let us introduce the following functions on the cotangent bundle $T^*M$ of the manifold $M$: 
\begin{align}
    T_i(x,p)=G^{AB}(x)\,p_A\,\partial_B\Omega_i(x)+\frac12\,\Omega_i(x)G^{AB}(x)\,p_Ap_B-\rho_i(x)\,.
\end{align}
We note the following properties. First, in a sufficiently small neighbourhood of a point on $\mathcal{C}$, the regularity conditions \eqref{regularity} imply that $T_i$ are functionally independent.
Second, we note that these functions are in involution with respect to the Poisson bracket $\{x^A,p_B\}=\delta^{A}_B$. Indeed, using \eqref{two-scales} and its differential consequence (see, e.g., \cite{BEG})
\begin{align}\label{rho-prolong}
\partial_A\rho_i=P_A{}^B\partial_B\Omega_i\,,
\end{align}
one finds
\begin{align}
    \{T_i,T_j\}=(\rho_j+\frac{\Omega_j}{2}G^{AB}p_Ap_B)T_i-(\rho_i+ \frac{\Omega_i}{2}G^{AB}p_Ap_B)T_j\,.
\end{align}
Hence, the system $T_i=0$ is involutive and by the standard local solvability result (see e.g. \cite[Article~428C, p.~1624]{ItoEDM}), the system of equations \eqref{rho-fix} admits a local solution.

It remains to compute the commutator of the gradients. 
 To this end, consider the following consequence of the first line in \eqref{two-scales}:
\begin{align}
    \nabla^A\Omega_1\,\nabla_A\nabla_B\Omega_2=-\nabla^A\Omega_1\,(P_{AB}\Omega_2+\rho_2G_{AB}).
\end{align}
Subtracting the analogous expression with $1$ and $2$ interchanged, we obtain
\begin{align}
    [\,\mathrm{grad}_G\,\Omega_1,\mathrm{grad}_G\,\Omega_2\,]^B&=\nabla^A\Omega_1\,\nabla_A\nabla^B\Omega_2-\nabla^A\Omega_2\,\nabla_A\nabla^B\Omega_1\\&=\nabla^A\Omega_2\,(P_{A}{}^{B}\Omega_1+\rho_1\delta_{A}^B)-\nabla^A\Omega_1\,(P_{A}{}^{B}\Omega_2+\rho_2\delta_{A}^B)\,.
\end{align}
Using \eqref{rho-prolong}, we can rewrite the commutator as
\begin{align}\label{U-comm}
    [\,\mathrm{grad}_G\,\Omega_1,\mathrm{grad}_G\,\Omega_2\,]^B=\Omega_1\nabla^B\rho_2+\rho_1\nabla^B\Omega_2-\Omega_2\nabla^B\rho_1-\rho_2\nabla^B\Omega_1\,.
\end{align}
Hence it is zero in the gauge $\rho_i=0$.

%%%%%%%%%%%
\subsection{Proof of Proposition \bref{propgamma-with-mass}}\label{proofgamma-with-mass}
%%%%%%%%%%%

    Substituting \eqref{gamma-with-mass} into \eqref{Spi-tilde-G} and introducing the coordinate $v$ by $u=\frac{\rho}{1-\epsilon v\rho}$, we obtain
\begin{align}\label{G-with-mass}
    \tilde G=\frac{1}{(1-\epsilon v \rho)^2}(2dvd\rho+\epsilon\rho^2dv^2+\frac{\nu(\rho,x)}{\rho}d\rho^2+h_{mn}(\rho,x)dx^mdx^n)\,.
\end{align}
This metric is manifestly singular at $\rho=0$ if $\nu(0,x)\neq0$, i.e., for a non-vanishing mass aspect. Nevertheless, this singularity has a very special form: it is confined to the transverse component of the metric.”. As we shall show, this singularity can be removed by a change  of coordinates, which is singular at the boundary but regular in the interior. It is convenient to present the inverse transformation:
\begin{align}\label{change-mass}
\begin{split}    
    x^m&=y^m+\frac{\rho\log\rho}{2}N^m(y)\,,\\
    v&=\tilde v-\frac{1}{2}\nu(0,y)\log\rho-\frac{\rho}{8}\partial_m\nu(0,y)N^m(y)((\log\rho)^2+2\log\rho)\,,
    \end{split}
\end{align}
where $N^m(y):= h^{mn}(0,y)\partial_n\nu(0,y)$.
After this change of coordinates, we obtain
\begin{multline}\label{G-with-mass1}
\tilde{G}=\frac{1}{(1-\epsilon\rho v(\tilde v,\rho,y))^2}\big[2d\tilde vd\rho +\big((1+\log\rho)h_{mn}(\rho,x(\rho,y))N^m-\log\rho\partial_n\nu(0,y)\big)dy^nd\rho +\\\big(\frac{\nu(\rho,x(\rho,y))-\nu(0,y)}{\rho}-\frac{1}{4}N^m\partial_m\nu(0,y)((\log\rho)^2+4\log\rho+2)+\frac{\epsilon}{4}\nu(0,y)^2+\\\frac{1}{4}(\log\rho+1)^2h_{mn}(\rho,x(\rho,y))N^mN^n\big)d\rho^2+h_{mn}(\rho,x(\rho,y))dy^mdy^n+\mathcal{O}(\rho|\log\rho|^3)\big]\,,
\end{multline}
where $\mathcal{O}(\rho(\log\rho)^3)$ denotes terms bounded by $C\rho|\log\rho|^3$ as $\rho\to0$. It is straightforward to see that the limit of $\tilde G$ as $\rho\to0$ exists and is given by
\begin{multline}
    \tilde G|_{\rho=0}=2d\tilde vd\rho+\partial_n\nu(0,y)dy^nd\rho+h_{mn}(0,y)dy^mdy^n+\\(\partial_\rho\nu(0,y)+\frac{\epsilon}{4}\nu(0,y)^2-\frac{1}{4}h^{mn}(0,y)\partial_m\nu(0,y)\partial_n\nu(0,y))d\rho^2\,.
\end{multline}
It follows that the metric \eqref{G-with-mass1} together with $u=\dfrac{\rho}{1-\epsilon\rho v(\tilde v,\rho, y)}$ provides the desired extension of the solution that is continuous up to the boundary. Note that the change of coordinates \eqref{change-mass} reduces to the identity when $\nu(0,y)=0$.

%\bibliographystyle{utphys}
%\bibliography{temporary, HSmaster}
\providecommand{\href}[2]{#2}\begingroup\raggedright\endgroup

\end{document}